\documentclass[pra,twocolumn,showpacs,letterpaper,showpacs,superscriptaddress]{revtex4-1}
\usepackage{amsmath,braket,amssymb,graphicx,stackrel,color,hyperref}

\usepackage {graphicx}
\usepackage{float}

\definecolor{darkblue}{rgb}{0.,0.,0.5}
\definecolor{darkred}{rgb}{0.5,0.,0.}
\definecolor{darkgreen}{rgb}{0.,0.5,0.}

\hypersetup{pdftitle={ExactSolutionsKerrTwoPhotons},colorlinks=true,linkcolor=darkred,citecolor=darkblue,urlcolor=darkblue}

\newcommand{\NN}{\mathcal{N}}
\newcommand{\FF}{\mathcal{F}}
\newcommand{\HH}{\mathcal{H}}
\newcommand{\DD}{\mathcal{D}}
\newcommand{\aop}{\hat{a}}
\newcommand{\ad}{\hat{a}^\dagger}
\newcommand{\Tr}[1]{\mathrm{Tr}\!\left[#1\right]}
\newcommand{\ada}[2]{\hat{a}^{\dag\, #1} \hat{a}^{#2}}
\newcommand{\corr}[2]{\braket{\ada{#1}{#2}}}

\renewcommand{\Re}{{\rm Re}}
\renewcommand{\Im}{{\rm Im}}

\begin{document}

\title{Exact steady state of a Kerr resonator with one- and two-photon driving and~dissipation: Controllable Wigner-function multimodality and dissipative phase~transitions}

\author{Nicola Bartolo}
\email{nicola.bartolo@univ-paris-diderot.fr}
\author{Fabrizio Minganti}
\author{Wim Casteels}
\author{Cristiano Ciuti}
\email{cristiano.ciuti@univ-paris-diderot.fr}
\affiliation{Universit\'{e} Paris Diderot, Sorbonne Paris Cit\'{e}, Laboratoire Mat\'{e}riaux et Ph\'{e}nom\`{e}nes Quantiques, CNRS-UMR7162, 75013 Paris, France}

\begin{abstract}
We present exact results for the steady-state density matrix of a general class of driven-dissipative systems consisting of a nonlinear Kerr resonator in the presence of both coherent (one-photon) and parametric (two-photon) driving and dissipation. Thanks to the analytical solution, obtained via the complex $P$-representation formalism, we are able to explore any regime, including photon blockade, multiphoton resonant effects, and a mesoscopic regime with large photon density and quantum correlations. We show how the interplay between one- and two-photon driving provides a way to control the multimodality of the Wigner function in regimes where the semiclassical theory exhibits multistability. We also study the emergence of dissipative phase transitions in the thermodynamic limit of large photon numbers.
\end{abstract}

\date{\today}


\maketitle

\section{Introduction}

Recently, the possibility to realize strong photon-photon interactions boosted the study of many-body physics with light~\cite{CarusottoRMP13}.
Indeed, new experimental platforms, such as semiconductor microcavities~\cite{WeisbuchPRL92,DeveaudBOOK} and superconducting circuits~\cite{SchoelkopfNature08,YouNature11}, allow one to realize photonic resonators with relatively large nonlinearities.
This enables the achievement of new highly-interacting regimes which, for decades, were confined to the textbook study of quantum optics~\cite{WallsBOOK}.
In this framework, a new flourishing field is that of reservoir engineering, whose goal is the manipulation of the photon exchanges between a nonlinear resonator and the environment~\cite{PoyatosPRL96,VerstraeteNatPhys09,TanPRA13,LinNature13,ArenzJPB13,AsjadJPB14,RoyPRA15,LeghtasScience15}.
These techniques permit the realization of new models with nontrivial drive and dissipation.
In this context, exactly-solvable models are of particular interest.
The analytic solution allows one to explore the full range of system parameters rather than the limiting regimes of small or high photon densities.
The latter are respectively tackled through numerical techniques and semiclassical approximations.
An example of a solvable model is the driven-dissipative Kerr model, for which Drummond and Walls derived the steady-state solution via the complex $P$-representation~\cite{DrummondJPA80a}.
Beyond the single resonator case, analytic solutions proved to be very useful for an efficient implementation of Gutzwiller mean-field decoupling for arrays of coupled cavities~\cite{DiehlPRL10,TomadinPRA11,LeBoitePRL13,LeBoitePRA14,JinPRL13,JinPRA14,WilsonarXiv16}.
	
In the present work, we use the complex $P$-representation to provide an exact solution for the steady state of a general class of driven-dissipative nonlinear resonators.
More precisely, we consider a standard driven-dissipative Kerr model (subject to the usual coherent pumping and one-photon dissipation) driven by an additional parametric two-photon pump and subject to two-photon losses.
Recently, these processes have been engineered for superconducting resonators~\cite{LeghtasScience15}.
The growing interest towards such kind of models is motivated by the emergence of nonclassical metastable and steady states in their dynamics, such as mixtures of quasi-coherent states or photonic Schr\"odinger cats, which lead to multi-modal Wigner functions~\cite{VogelPRA89,KerckhoffOE11,MingantiSciRep16}.
The possibility to control and protect such states is promising for the implementation of quantum computation protocols~\cite{GilchristJOB04,OurjoumtsevScience06,MirrahimiNJP14,GotoPRA16,PuriarXiv16}.
The exact solution derived in this work allows one to explore the quantum properties of the steady state beyond the semiclassical approximation, capturing the emergence of dissipative phase transitions~\cite{CarmichaelPRX15,CasteelsarXiv16}.
Furthermore, the exact solution paves the way to a numerically-efficient exploration of resonator lattices through Gutzwiller decoupling.

The paper is organized as follows.
We start in Sec.~\ref{Sec:System} by introducing the model.
Than, in Sec.~\ref{Sec:Solution}, we exploit the formalism of the complex $P$-representation to derive the exact solution for the steady-state of the considered model.
Section~\ref{Sec:Study} is devoted to the study of the steady-state properties.
We compare semiclassical and quantum solutions in Sec.~\ref{subsec:semiclassical}.
In Sec.~\ref{Subsec:PhaseTransitions}, we show the emergence of dissipative phase transitions in the thermodynamic limit of large excitation numbers.
Finally, we present conclusions and perspectives in Sec.~\ref{sec:conclusion}.


\begin{figure}[t!]
	\includegraphics[width=0.48\textwidth]{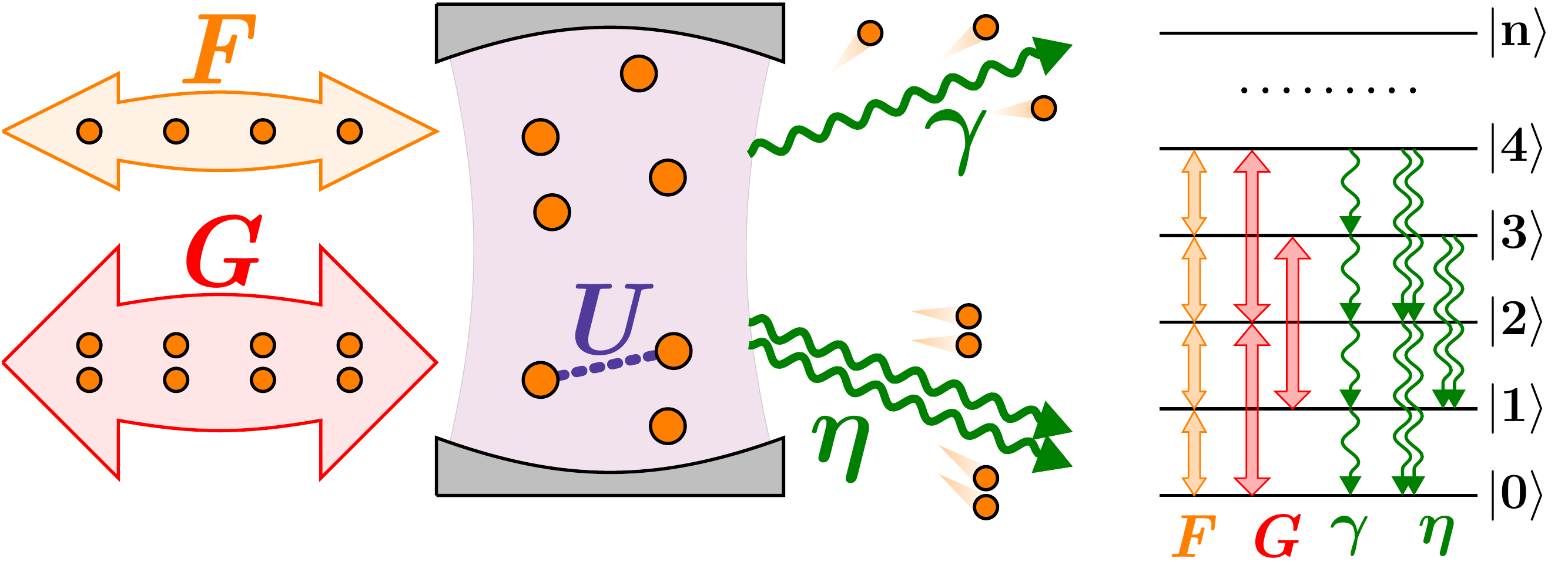}
	\caption{A sketch of the considered class of systems.
	The picture represents a photon resonator subject to one-photon losses with rate $\gamma$, and coherently driven by a one-photon pump of amplitude $F$.
	The resonator is also subject to a coherent two-photon driving of amplitude $G$ and two-photon losses with rate $\eta$.
	The strength of the photon-photon interaction is quantified by $U$.
	On the right, we sketch the effects of these physical processes on the Fock (number) states~$\vert n \rangle$.}
\label{fig:syst_ph}

\end{figure}

\section{Nonlinear resonator including two-photon processes}\label{Sec:System}
Let us introduce the general model of a driven-dissipative Kerr nonlinear resonator with two-photon drive and dissipation which is sketched in Fig. \ref{fig:syst_ph}. In the the Hamiltonian, $\omega_c$ is the cavity-mode frequency and $U$ the strength of the Kerr nonlinearity, quantifying the photon-photon interaction.
In the absence of pumping we get ($\hbar=1$)
\begin{equation}
\hat{\HH}_0=\omega_c\,\ad\aop+\frac{U}{2}\ad\ad\aop\aop,
\end{equation}
where $\hat{a}$ and $\hat{a}^\dagger$ are, respectively, the annihilation and creation operator for photons inside the resonator.
A coherent drive with amplitude $F$ and frequency $\omega_p$ can be described by
\begin{equation}\label{Eq:H1ph}
\hat{\HH}_{\rm1ph}=F\,e^{- i  \omega_p t}\ad+F^*\,e^{ i  \omega_p t}\aop.
\end{equation}
From now on we will denote this mechanism as one-photon pumping.
Similarly, a parametric process coherently adding photons pairwise is described by
\begin{equation}\label{Eq:H2ph}
\hat{\HH}_{\rm2ph}=\frac{G}{2}\,e^{-i \omega_2t}\ad\ad+\frac{G^*}{2}\,e^{i \omega_2t}\aop\aop,
\end{equation}
where $G$ is the pump amplitude and $\omega_2$ its frequency.
Such a two-photon pumping mechanism can be obtained by engineering the exchange of photons between the cavity and the environment.
Recently, this has been realized by coupling two superconducting resonators via a Josephson junction~\cite{LeghtasScience15}.
In order to get a time-independent Hamiltonian, we consider $\omega_2=2\omega_p$.
Hence, we use the unitary transformation $\hat{\mathcal{U}}=e^{- i \omega_pt\ad\aop}$, which removes the time-dependence from the Hamiltonian.
This allows us to describe the system in the reference frame rotating at the coherent pump frequency $\omega_p$.
The full Hamiltonian, hence, becomes
\begin{align}\label{Eq:HCompleteOur}
\hat{\HH}=&-\Delta\ad\aop+\frac{U}{2}\ad\ad\aop\aop
\nonumber\\
&+F\ad+F^*\aop+\frac{G}{2}\ad\ad+\frac{G^*}{2}\aop\aop,
\end{align}
where $\Delta=\omega_p-\omega_c$ is the pump-cavity detuning.
For the considered system, photon losses are typically appreciable and can not be neglected~\cite{HarocheBOOK}.
The Markov-Born approximation gives an excellent description of these losses in terms of a Lindblad dissipation super-operator $\mathcal{D}(\hat{\mathcal{C}})$ of the form ~\cite{HarocheBOOK, CarmichaelBOOK}
\begin{equation}
	\mathcal{D}(\hat{\mathcal{C}})\,\hat{\rho}=
	2\,\hat{\mathcal{C}}\,\hat{\rho}\,\hat{\mathcal{C}}^\dagger
	-\hat{\mathcal{C}}^\dagger \hat{\mathcal{C}}\,\hat{\rho}
	-\hat{\rho}\,\hat{\mathcal{C}}^\dagger\hat{\mathcal{C}},
\end{equation}
where $\hat{\mathcal{C}}$ is the quantum jump operator corresponding to the specific dissipation process.
Usually, photons are lost individually to the environment and the jump operator is the annihilation operator $\aop$~\cite{HarocheBOOK}. In addition, we also consider two-photon losses, which naturally emerge together with the engineered two-photon pumping~\cite{LeghtasScience15}.
These losses are included through the jump operator $\aop^2$.
The resulting Lindblad master equation describing the evolution of the the system density matrix $\hat{\rho}$ is
\begin{equation}\label{Eq:MasterEquationOur}
i \frac{\partial\hat{\rho}}{\partial t}=
\left[\hat{\HH},\hat{\rho}\right]
+ i \frac{\gamma}{2}\,\DD(\aop)\,\hat{\rho}
+ i \frac{\eta}{2}\,\DD(\aop^2)\,\hat{\rho},
\end{equation}
where $\gamma$ and $\eta$ are, respectively, the one- and two-photon dissipation rates and $\hat{\mathcal{H}}$ is the one given in Eq.~\eqref{Eq:HCompleteOur}.


\section{$P$-representation and exact solution for the steady state}\label{Sec:Solution}
The steady-state properties are of central interest in the context of out-of-equilibrium quantum systems. These properties are encoded in the steady-state density matrix, which is the solution of Eq.~\eqref{Eq:MasterEquationOur} for $\partial_t\hat{\rho}=0$.
To this purpose, we consider the $P$-representation of the density matrix, i.e. we decompose $\hat{\rho}$ using the over-complete basis of coherent states $\ket{\alpha}$, such that $\aop\ket{\alpha}=\alpha\ket{\alpha}$.
We use the complex $P$-representation $P(\alpha,\beta)$~\cite{DrummondJPA80b}, which is defined by
\begin{equation}\label{Eq:PRepresentation}
\hat{\rho}=\int_\mathcal{C}\!\!d\alpha \int_{\mathcal{C}'}\!\!d\beta\,
\frac{\ket{\alpha}\bra{\beta^*}}{\braket{\beta^*|\alpha}}\,
P(\alpha,\beta),
\end{equation}
where the closed integration contours $\mathcal{C}$ and $\mathcal{C}'$ must be carefully chosen to encircle all the singularities of the function $P(\alpha,\beta)$.
Once the definition~\eqref{Eq:PRepresentation} is inserted in Eq.~\eqref{Eq:MasterEquationOur}, the action of the annihilation and creation operators on the projector $\ket{\alpha}\!\bra{\beta^*}$ allows one to map the master equation for $\hat{\rho}$ into a complex Fokker-Planck equation for $P(\alpha,\beta)$. Further details on this procedure are presented in appendix~\ref{App:Solution}.
For the case $G=0$, the complex $P$-representation solution for the steady state was derived by Drummond and Walls~\cite{DrummondJPA80a}, and is given by
\begin{equation}\label{Eq:PDrummondWalls}
P_{\rm ss}(\alpha,\beta)\propto e^{2\alpha\beta}
\frac{e^{-2f/\alpha}}{\alpha^{2+2c}}
\frac{e^{-2f^*\!/\beta}}{\beta^{2+2c^*}}.
\end{equation}
In Eq.~\eqref{Eq:PDrummondWalls}, the system parameters are resumed by the dimensionless quantities \mbox{$c=(\Delta+i\gamma/2)/(U-i\eta)$} and \mbox{$f=F/(U-i\eta)$}.
For the general case corresponding to the master equation~\eqref{Eq:MasterEquationOur}, we find
\begin{equation}\label{Eq:PRepOur}
\begin{split}
P_{\rm ss}(\alpha,&\beta)=\frac{
	e^{2\alpha\beta}}{\NN}
\frac{1}{\left(\alpha^2+g\right)^{1+c}}
\exp\left[-\frac{2f}{\sqrt{g}}\arctan\left(\frac{\sqrt{g}}{\alpha}\right)\right]\\
&\times\,\frac{1}{\left(\beta^2+g^*\right)^{1+c^*}}
\exp\left[-\frac{2f^*}{\sqrt{g^*}}\arctan\left(\frac{\sqrt{g^*}}{\beta}\right)\right].
\end{split}
\end{equation}
All details on the derivation of Eq.~\eqref{Eq:PRepOur} are given in appendix~\ref{App:Solution}.
In Eq.~\eqref{Eq:PRepOur} we introduced the dimensionless parameter \mbox{$g=G/(U-i\eta)$}.
We stress that in the limit $g\to0$ Eq.~\eqref{Eq:PRepOur} reduces to Eq.~\eqref{Eq:PDrummondWalls}, as expected. We note that some particular cases have been considered in~\cite{KryuchkyanOC96,MeaneyEPJQT14,ElliottarXiv16}.

The normalization factor $\NN$ in Eq.~\eqref{Eq:PRepOur} ensures that $\Tr{\hat{\rho}}=1$.
By imposing this condition we get
\begin{align}
\NN=&
\int_\mathcal{C}\!\!d\alpha \int_{\mathcal{C}'}\!\!d\beta\,\,
e^{2\alpha\beta} \frac{1}{\left(\alpha^2+g\right)^{1+c}}
\frac{1}{\left(\beta^2+g^*\right)^{1+c^*}}
\nonumber\\
\times&\exp\left[-\frac{2f}{\sqrt{g}}\arctan\left(\frac{\sqrt{g}}{\alpha}\right)
-\frac{2f^*}{\sqrt{g^*}}\arctan\left(\frac{\sqrt{g^*}}{\beta}\right)\right].
\end{align}
One can Taylor-expand $e^{2\alpha\beta}$ and swap the resulting sum with the integral.
The two contour integrals over $\alpha$ and $\beta$ thus decouple, leading to
\begin{align}\label{Eq:Normalization}
\NN=\sum_{m=0}^{\infty} \frac{2^m}{m!}\, \left|\FF_m\left(f,g,c\right)\right|^2,
\end{align}
where we introduced
\begin{equation}\label{Eq:DefinitionFFormal}
\FF_m\left(f,g,c\right)=\!\!\int_\mathcal{\!C} 
\frac{\alpha^m\, d\alpha}{\left(g+\alpha^2\right)^{1+c}}\,
\exp\left[
-\frac{2f}{\sqrt{g}}\,\arctan\left(\frac{\sqrt{g}}{\alpha}\right)
\right]\!.
\end{equation}
Note that, $\FF_m\left[f^*,g^*,c^*\right]=\FF_m^*\left[f,g,c\right]$.
Performing the integral in Eq.~\eqref{Eq:DefinitionFFormal} requires an appropriate choice
of the contour $\mathcal{C}$.
In the present case, we used the Pochhammer path (more details are given in appendix~\ref{App:Solution}), which gives
\begin{equation}
\label{Eq:FormulaFcomplete}
\FF_m\left(f,g,c\right)=\left(i\sqrt{g}\right)^m
\,_2 F_1\left(-m,-c-i\,f/\sqrt{g};-2c;2\right),
\end{equation}
where $\,_2 F_1$ is the Gauss hypergeometric function~\cite{BaileyBOOK}.


\subsection{Exact results for steady-state quantities}\label{subsec:ssvalues}

The steady-state quantities can be expressed in terms of the $\FF_m$ functions~\eqref{Eq:FormulaFcomplete}.
Let us consider the correlation functions
\begin{align}\label{Eq:CorrijFormal}
\corr{i}{j}&=\Tr{\ada{i}{j}\,\hat{\rho}}
\nonumber\\&=
\!\!\int_\mathcal{C}\!\!d\alpha \int_{\mathcal{C}'}\!\!d\beta\,\,
\frac{P(\alpha,\beta)}{\braket{\beta^*|\alpha}}\,\,
\Tr{\ada{i}{j}\ket{\alpha}\bra{\beta^*}}.
\end{align}
Since $\Tr{\ada{i}{j} \ket{\alpha}\bra{\beta^*}}=\alpha^j\beta^i\braket{\beta^*|\alpha}$, we have
\begin{equation}
\label{Eq:CorrijGeneral}
\corr{i}{j}=\frac{1}{\NN}\,\sum_{m=0}^{\infty} \frac{2^m}{m!}\, \FF_{m+j}\left[f,g,c\right]\, \FF_{m+i}^*\left[f,g,c\right].
\end{equation}
Similarly, one can derive the matrix elements of the steady-state density matrix $\hat{\rho}_{\rm ss}$ in the Fock basis:
\begin{align}\label{Eq:RhopqGeneral}
&\braket{p|\hat{\rho}_{\rm ss}|q}=
\!\!\int_\mathcal{C}\!\!d\alpha \int_{\mathcal{C}'}\!\!d\beta\,\,
\frac{P(\alpha,\beta)}{\braket{\beta^*|\alpha}}
\frac{\alpha^p\beta^q}{\sqrt{p!q!}}
\nonumber\\
&\quad=\frac{1}{\NN\,\sqrt{p!q!}} \sum_{m=0}^{\infty} \frac{1}{m!}
\FF_{m+p}\left[f,g,c\right]\, \FF_{m+q}^*\left[f,g,c\right].
\end{align}

An useful tool to visualize the properties of the steady state is the Wigner function~\cite{WignerPR32}.
Given a density matrix $\hat{\rho}$, the corresponding Wigner function $W(z)$ is a real-valued function of the complex variable $z$, defined as~\cite{CahillPR69}
\begin{equation}
W(z)=\frac{2}{\pi}\,\Tr{\hat{D}_z\, e^{i\pi \ad\aop}\, \hat{D}_z^\dagger\, \hat{\rho}},
\end{equation}
with $\hat{D}_z=e^{z \ad- z^* \aop}$ the displacement operator.
Substituting $\hat{\rho}$ with its $P$-representation, the crucial quantity to evaluate is $\Tr{\hat{D}_z\, e^{i\pi \ad\aop}\, \hat{D}_z^\dagger\ket{\alpha}\bra{\beta^*}}$.
Using the identity $\hat{D}_z^\dagger\,\hat{D}_\alpha=e^{(\alpha z^*\!-z\alpha^*)/2} \hat{D}_{\alpha-z}$, after lengthy but straightforward calculations, one gets
\begin{align}
\Tr{\hat{D}_z\, e^{i\pi \ad\aop}\, \hat{D}_z^\dagger\ket{\alpha}\bra{\beta^*}}
=\braket{\beta^*|\alpha}\frac{e^{2\alpha z^*} e^{2\beta z}}{e^{2\alpha\beta} e^{2|z|^2}}.
\end{align}
The Wigner function can thus be written as
\begin{equation}
W(z)=\frac{2\,e^{-2|z|^2}}{\pi}
\int_\mathcal{C}\!\!d\alpha \int_{\mathcal{C}'}\!\!d\beta\,\,
\frac{P(\alpha,\beta)}{e^{2\alpha\beta}}\,
e^{2\alpha z^*} e^{2\beta z}.
\end{equation}
This time, the integrals over $\alpha$ and $\beta$ are already independent.
By Taylor expanding the exponentials, we find that
\begin{equation}
\label{Eq:WignerGeneral}
W(z)=\frac{2}{\pi\,\NN} \left|
\sum_{m=0}^{\infty} \frac{(2z^*)^m}{m!}\, \FF_m\left[f,g,c\right]
\right|^2 e^{-2|z|^2}.
\end{equation}
Therefore, the Wigner function given in Eq.~\eqref{Eq:WignerGeneral} is real and positive over the whole complex plane for \emph{any} choice of the system parameters.

We point out that Eqs.~\eqref{Eq:Normalization}, \eqref{Eq:CorrijGeneral}, \eqref{Eq:RhopqGeneral}, and~\eqref{Eq:WignerGeneral}, together with the definition of $\FF_m$ given in Eq.~\eqref{Eq:FormulaFcomplete}, summarizes the exact analytic results of this work.
For sake of completeness, in the case $g=0$, the definition~\eqref{Eq:FormulaFcomplete} can be reduced to \mbox{$\FF_m(f,0,c)=(-2f)^m/\Gamma(m-2c)$}~\cite{DrummondJPA80a}.
Although the exact results presented here apply for the general case of complex $F$ and $G$, in what follows, unless differently specified, we will take them as real parameters.

\subsection{Benchmarking in the low-density regime}

\begin{figure}[t]
	\includegraphics[width=0.42\textwidth]{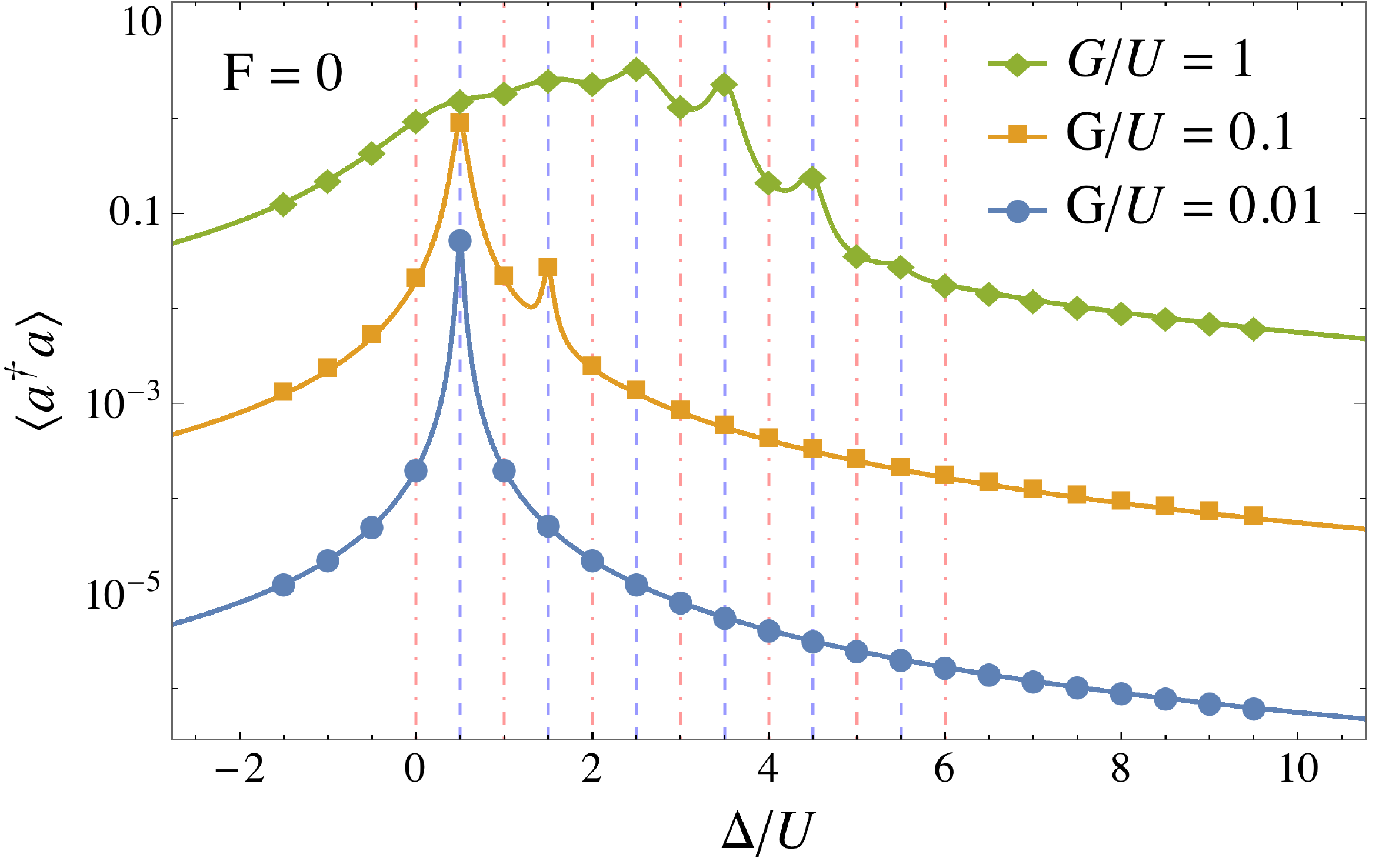}\vspace{0.3cm}
	\includegraphics[width=0.42\textwidth]{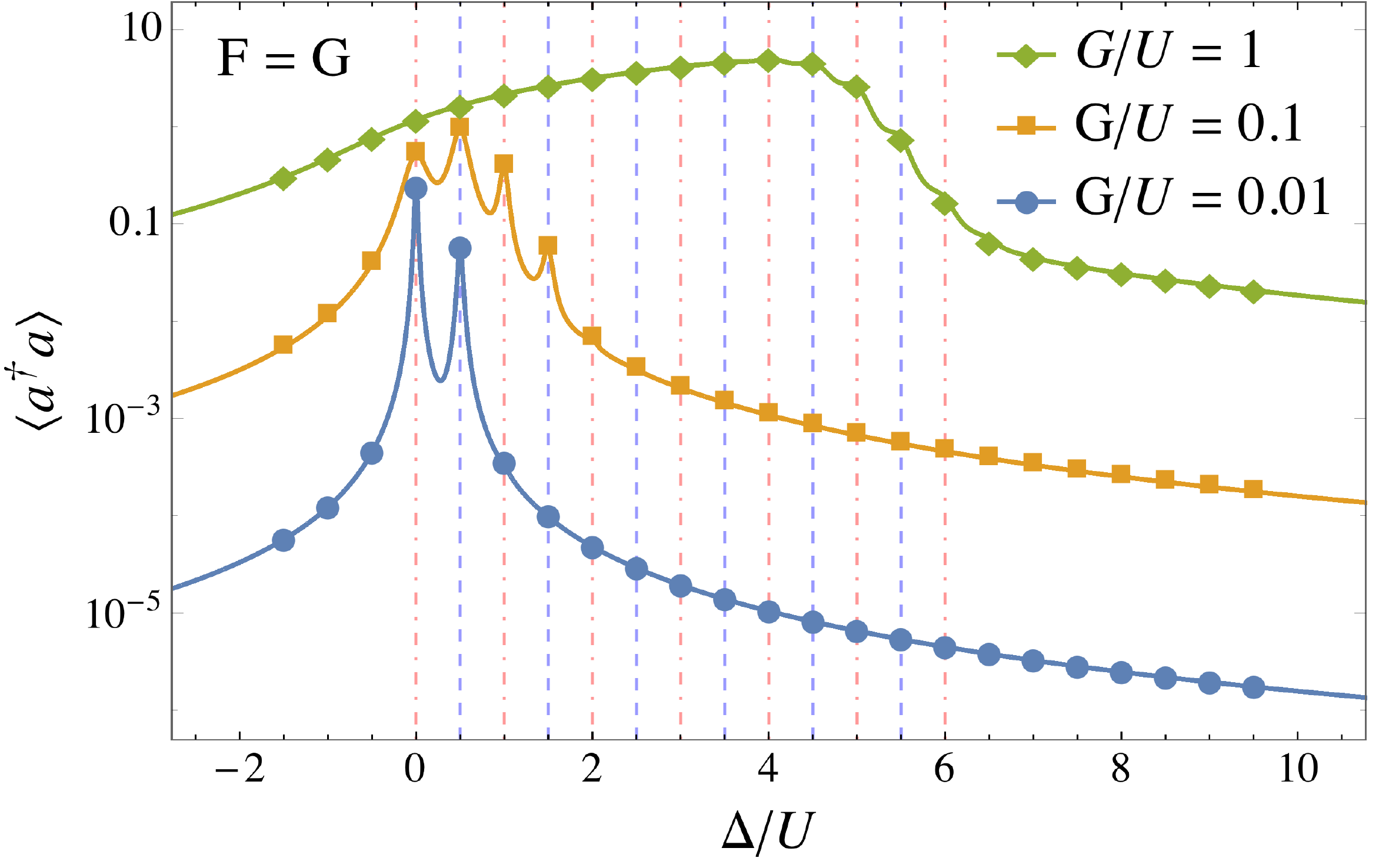}
	\caption{(Color online)
		Mean photon number $\braket{\ada{}{}}$ as a function of the pump-cavity detuning $\Delta$ normalized by the photon-photon interaction strength $U$.
		Different curves and data sets correspond to different pump intensities (cf. legend).
		Solid lines represent the analytic solution while the points are the numerical results obtained by diagonalization of the Liouvillian superoperator of the master equation on a truncated Fock basis.
		Top: results in the absence of one-photon pumping, i.e. $F=0$ [Eq.~\eqref{Eq:CorrijNoF} for $i=j=1$].
		Bottom: results in the presence of both one- and two-photon pumping with $F=G$ [Eq.~\eqref{Eq:CorrijGeneral}].
		In both panels, vertical dot-dashed red (dashed blue) lines mark the position of odd (even) photonic resonances.
		One- and two-photon dissipation rates were set to $\gamma=\eta=0.03U$.}
	\label{Fig:Numerics}
\end{figure}

Before exploiting the analytic solution, note that the results summarized in   Eqs.~\eqref{Eq:Normalization}, \eqref{Eq:CorrijGeneral}, \eqref{Eq:RhopqGeneral}, and~\eqref{Eq:WignerGeneral}, contain infinite sums of $\mathcal{F}_m$ functions.
In the special cases $G=0$ (studied in~\cite{DrummondJPA80a}) or $F=0$ (cf. appendix~\ref{subsec:closedForms}) such sums can be analytically computed, resulting in combinations of hypergeometric functions.
For the general case of finite one- and two-photon pumping (i.e., $F,G\neq0$), the series can be computed with arbitrary precision (see appendix~\ref{Subsec:Convergence} for further details).

In order to benchmark the analytic solution with numerical approaches, we study it in the low-density regime.
The two panels of Fig.~\ref{Fig:Numerics} show the results obtained in the presence of only two-photon pumping (top) and for both one- and two-photon driving (bottom).
The agreement with numerics is excellent, thus corroborating the validity of the analytic solution.
The parameters have been chosen to clearly visualize the photonic resonances, which are expected when the energy of $n$ pump photons is equal to that of $n$ photons inside the resonator.
Thus, beside the one-photon resonance occurring for $\Delta=0$, the multi-photon resonances arise when $\Delta/U=(n-1)/2$.
For $F=0$ only resonances with an even number of photons appear, while all of them are observed in the presence of a one-photon pumping.
The resonances progressively merge in a continuum by increasing the pump intensities.
In the high-density regime this behavior triggers a dissipative phase transition~\cite{CarmichaelPRX15}, discussed below in Sec.~\ref{Subsec:PhaseTransitions}.

\section{Properties of the steady state}\label{Sec:Study}

\begin{figure}[t]
	\includegraphics[width=0.48\textwidth]{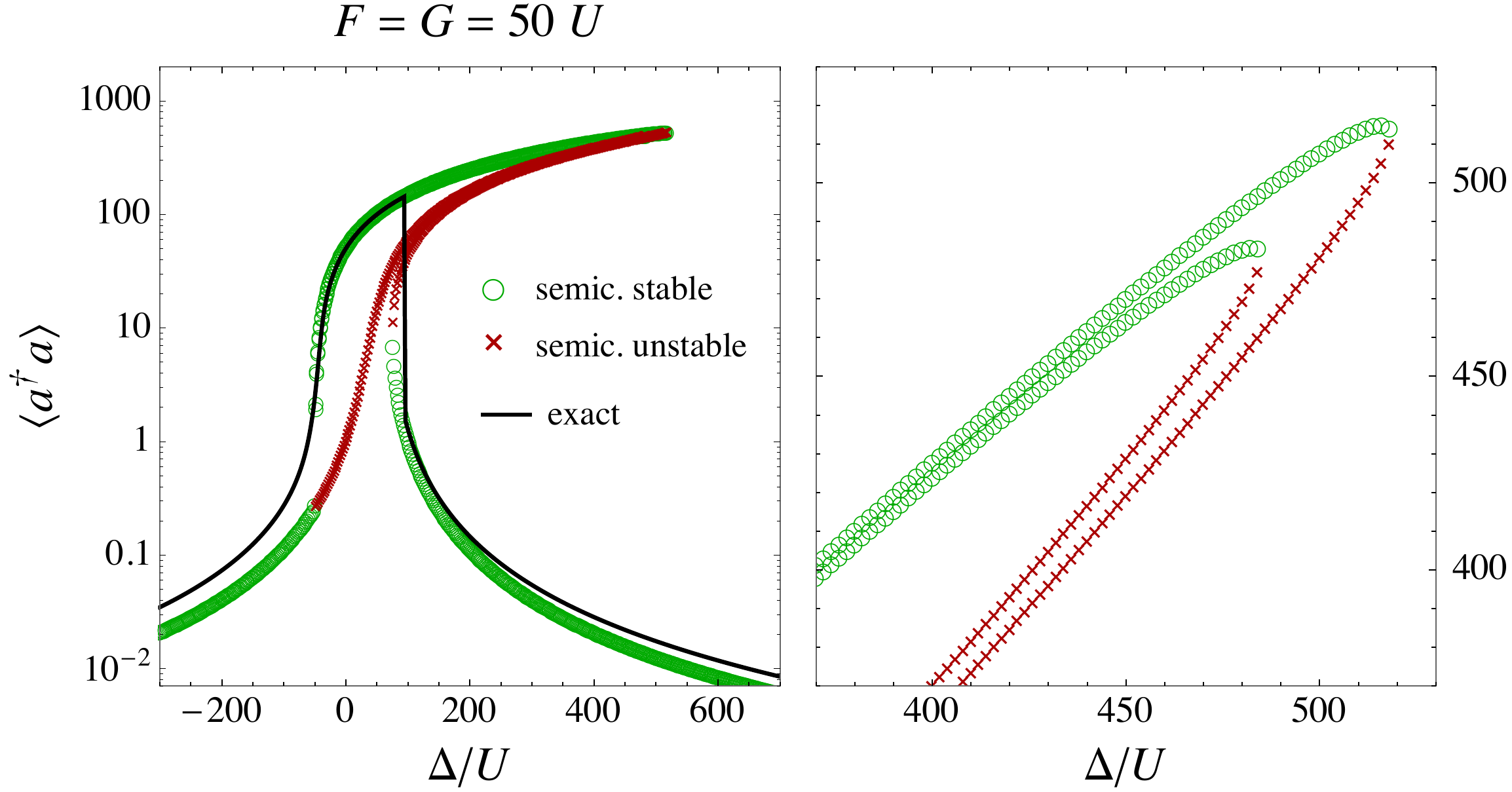}
	\caption{(Color online)
		Left: mean steady-state photon number $\corr{}{}$ as a function of the dimensionless detuning parameter $\Delta/U$.
		The green circles (red crosses) mark the stable (unstable) semiclassical steady-state solutions.
		The black line is the analytic solution given by Eq.~\eqref{Eq:CorrijGeneral} ($i=j=1$).
		One- and two-photon dissipation rates were set to $\gamma=\eta=0.1U$.
		Right: zoom-in on the region in which the almost-degenerate high-density semiclassical solutions get unstable.}
	\label{Fig:GP}
\end{figure}

\begin{figure*}[t]
	\includegraphics[width=0.85\textwidth]{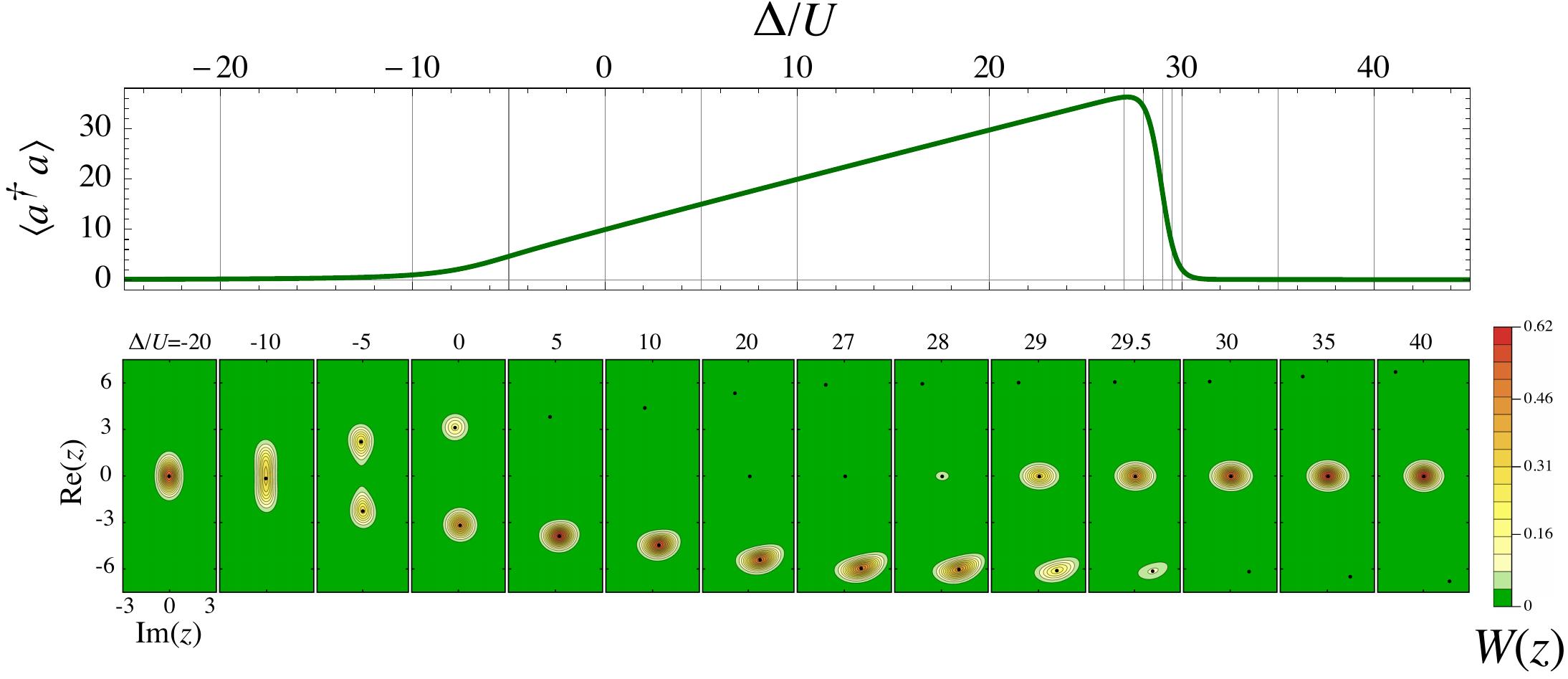}
	\caption{(Color online)
		Top: steady-state photon number $\corr{}{}$ as a function of the dimensionless detuning parameter $\Delta/U$ for $F=U$, $G=10U$, $\gamma=\eta=0.1U$.
		The results have been obtained through the exact solution~\eqref{Eq:CorrijGeneral} for $i=j=1$.
		The vertical grid lines mark the values of $\Delta/U$ for which we evaluated the steady-state Wigner function (cf. bottom panels).
		Bottom: steady-state Wigner functions $W(z)$ calculated according to Eq.~\eqref{Eq:WignerGeneral} for the same parameters as in the top panel and for different values of $\Delta/U$ (see frame labels).
		The black dots mark the position of the corresponding stable semiclassical solutions.}
	\label{Fig:Wigners}
\end{figure*}

The exact analytic solution of the Lindblad equation~\eqref{Eq:MasterEquationOur}, derived and benchmarked in Sec.~\ref{Sec:Solution}, allows us to compute the average steady-state quantities of the considered system in any physical regime, from low- to high-density phases, passing through the nontrivial mesoscopic regime.
In this section, we investigate how the properties of the steady state evolve through these different regimes.

\subsection{Quantum \emph{vs} semiclassical behavior}\label{subsec:semiclassical}

When the resonator has a small population, the solution of the master equation~\eqref{Eq:MasterEquationOur} can be obtained numerically.
For instance, by integrating the master equation on a truncated Fock basis or by diagonalizing the Liouvillian super-operator~\cite{CasteelsPRA16}.
On the other hand, for high photon number the cavity field can be typically approximated by a coherent state $\ket{\alpha}$.
Thus, the master equation reduces to a differential equation for the complex amplitude $\alpha$.
This corresponds to the so-called semiclassical approximation~\cite{CarusottoRMP13}.
In this case, however, all quantum correlations are neglected, which makes our exact analytic solution a precious tool.
The differential equation for the complex amplitude $\alpha$ can be easily derived from $\partial_t\braket{\hat{a}}=\Tr{\hat{a}\,\partial_t \rho}$, by assuming $\hat{\rho}=\ket{\alpha}\!\bra{\alpha}$, namely:
\begin{equation}\label{Eq:GP}
	\partial_t\alpha=(i\Delta-\gamma/2)\alpha -iF -iG\alpha^* -(iU+\eta)\alpha^*\alpha^2.
\end{equation}
Note that the latter equation is coupled to the one for the conjugate variable $\alpha^*$.
Solving for the steady state $\partial_t\alpha,\partial_t\alpha^*=0$ one can get, depending on the system parameters, up to five solutions, of which at most three are dynamically stable~\cite{MeaneyEPJQT14,ElliottarXiv16}.

In Fig.~\ref{Fig:GP} we show the semiclassical prediction for the mean photon number according to the semiclassical analysis.
For large and negative detuning, Eq.~\eqref{Eq:GP} predicts a single low-density steady-state solution.
By increasing $\Delta$, the low-density solution gets unstable and two high-density ones emerge.
The corresponding values of $\corr{}{}$ are almost equal, but the phases of their complex amplitudes differ approximatively by $\pi$.
Eventually, a third low-density stable solution appears, coexisting with the two high-density ones until a parameter-dependent threshold is reached (see zoom-in panel in Fig.~\ref{Fig:GP}).
Then, only the low-density stable state is present.
By comparing these results with the exact one given by Eq.~\eqref{Eq:CorrijGeneral} (also plotted in Fig.~\ref{Fig:GP}), we note that the multi-stable behavior does not appear in the analytic solution.
We point out that the quantum solution is unique, while the semiclassical approach gives multiple dynamically stable solutions.
However, the exact and unique quantum solution can display a multimodal mixed-state behavior.

The presence of one (or more) semiclassical solution(s) in the steady state can be visualized by the Wigner function $W(z)$, whose analytic expression is in Eq.~\eqref{Eq:WignerGeneral}.
The case $F=0$ has already been discussed in~\cite{MeaneyEPJQT14}, in particular the evolution of $W(z)$ across the density drop.
We present, in Fig.~\ref{Fig:Wigners}, the results for the general case $F,G\neq0$.
In the multiple-solution regime, even for $F/G\ll1$, the one-photon driving prevents the system from being in a balanced mixture of coherent states, which is the case for $F=0$~\cite{MeaneyEPJQT14,MirrahimiNJP14,LeghtasScience15,MingantiSciRep16,ElliottarXiv16}.
By looking at the bottom panel of Fig.~\ref{Fig:Wigners}, one notes that a bimodal Wigner function only exists nearby the transitions from low- to high-density regimes.
Elsewhere, $W(z)$ always exhibits a single peak.
In the low-density regimes, we recover a squeezed-vacuum steady state as the one observed for $F=0$~\cite{MeaneyEPJQT14,ElliottarXiv16}.
This squeezing of the state can be seen by looking at the elongated elliptic shape of the corresponding Wigner function in the bottom panels of Fig.~\ref{Fig:Wigners}.

\begin{figure}[t]
	\includegraphics[width=0.48\textwidth]{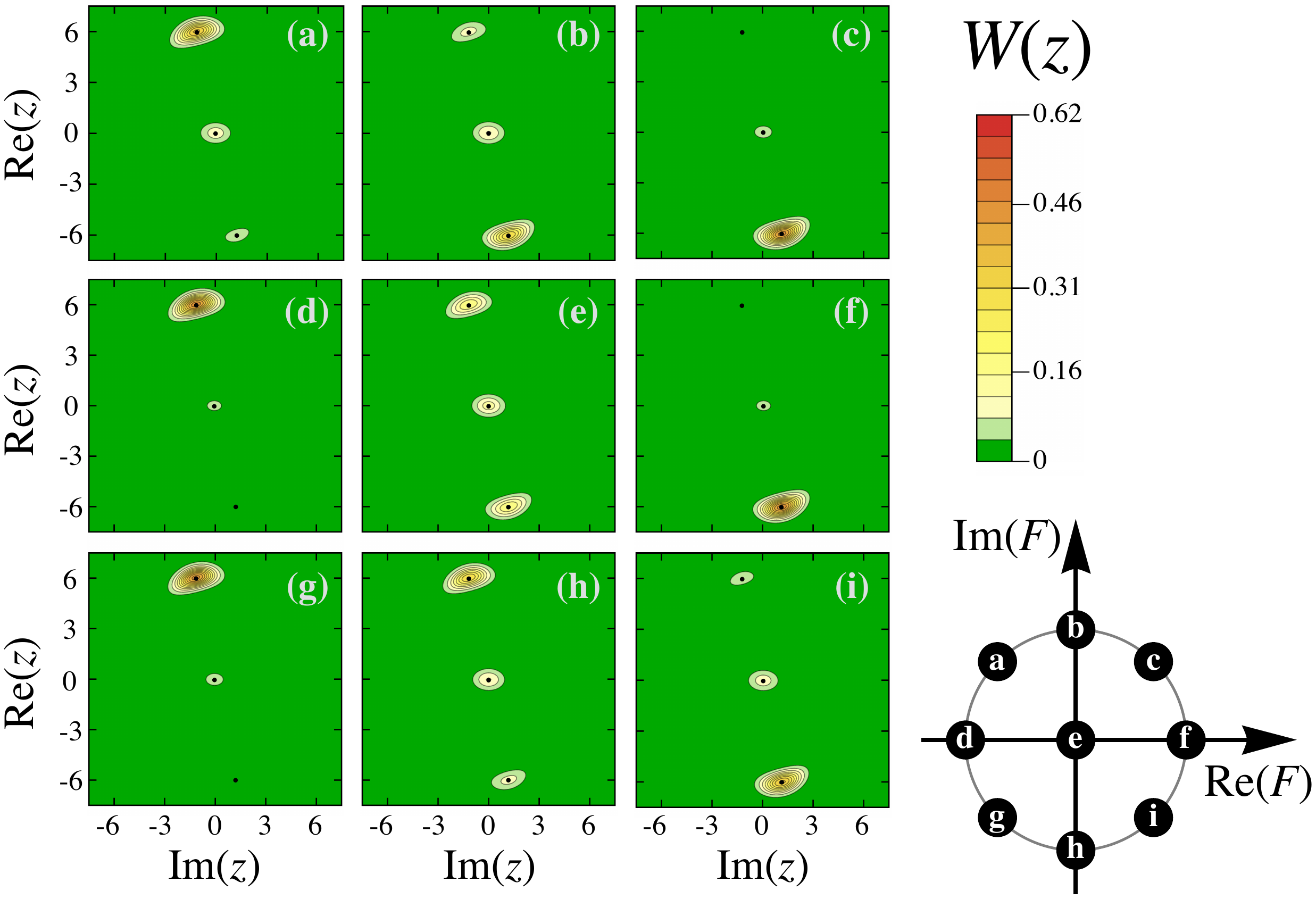}
\caption{(Color online)
	Steady-state Wigner functions $W(z)$ calculated according to Eq.~\eqref{Eq:WignerGeneral} for $\Delta=28U$, $G=10U$, $\gamma=\eta=0.1U$ and for different complex values of $F$.
	For panel~(e) we took $F=0$.
	In the others, $F/U=e^{i\phi}$ and the phase $\phi$ changes as sketched in the bottom-right scheme.}\label{Fig:PhaseF}
\end{figure}

Remarkably, as shown in Fig.~\ref{Fig:PhaseF}, the dominant peak in the multi-modal Wigner function is selected by the relative phase of $F$ and $G$.
For this analysis, we took the same parameters as in Fig.~\ref{Fig:Wigners}, setting the detuning around the threshold value.
In the outer panels we have varied the relative phase $\phi=\arg(F/G)$, changing the relative weight of the Wigner-function peaks.
The central panel~(e) shows the case $F=0$, for which the three peaks have comparable heights.
This property can be a valuable tool for the control of two-photon driven resonators for quantum computation based on quasi-orthogonal coherent states~\cite{GotoPRA16,PuriarXiv16}.
Indeed the relative phase $\phi$ could be experimentally controlled and adjusted at will.
In this direction, it is worth stressing that expression~\eqref{Eq:WignerGeneral} allows to predict precisely the shape of the multi-modal Wigner function even in highly populated regimes, where a numerical approach would be extremely demanding.

\subsection{Emergence of dissipative phase transitions}\label{Subsec:PhaseTransitions}

\begin{figure}[t]
	\includegraphics[width=0.45\textwidth]{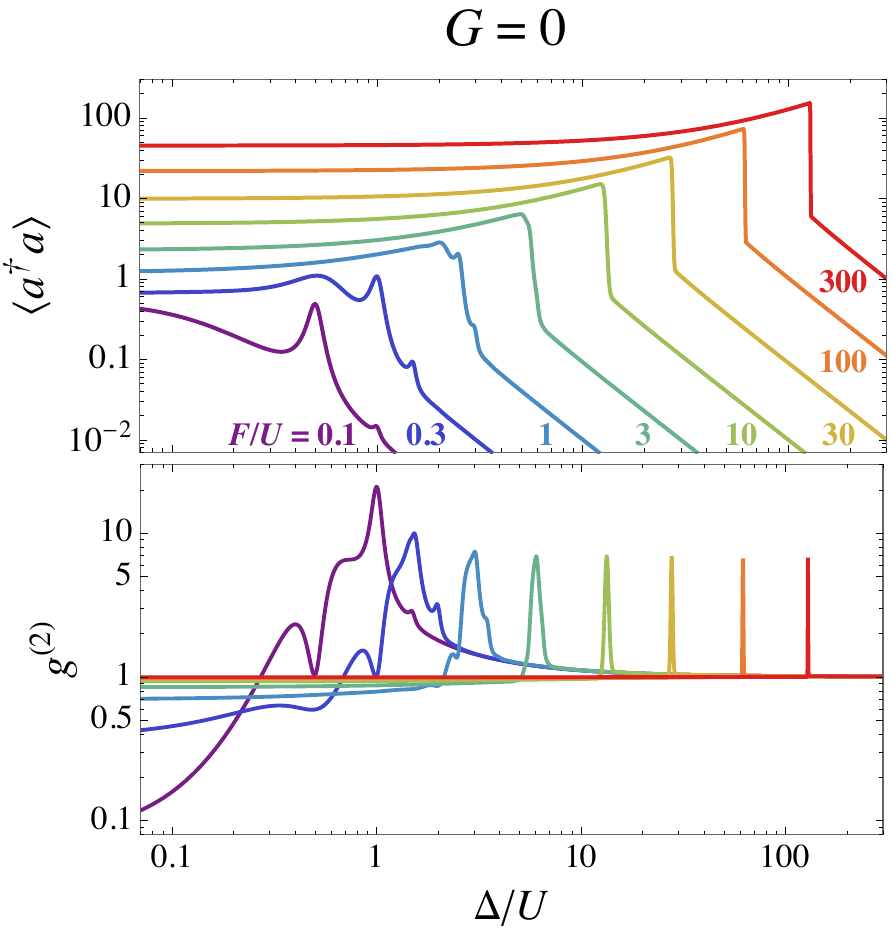}
	\caption{(Color online)
		Mean photon number $\braket{\ada{}{}}$ (top panel) and normalized second-order correlation function $g^{(2)}$ (bottom panel) as a function of the pump-cavity detuning $\Delta$ normalized by the photon-photon interaction strength $U$ for a resonator subject only to one-photon coherent driving ($G=0$, $F\neq0$).
		Different curves (and colors) correspond to different pumping intensities $F/U$, varied between 0.1 and 300, as indicated beside each curve in the top panel.
		One- and two-photon dissipation rates were set to $\gamma=\eta=0.03U$.}
	\label{Fig:BreakdownG0}
\end{figure}

In this section, we show how our analytic solution allows to capture the steady-state properties of dissipative phase transitions in the thermodynamic limit.
The latter, in the present context, is defined as the regime in which $\corr{}{}\to+\infty$~\cite{CarmichaelPRX15,CasteelsarXiv16}.
Let us start by considering the case in which the resonator is subject only to a coherent drive (i.e., $G=0$).
In the top panel of Fig.~\ref{Fig:BreakdownG0} we show the evolution of the mean photon density $\corr{}{}$ as a function of the detuning for different driving amplitudes $F$.
For a small drive amplitude $F\lesssim U$, the photon number shows well-resolved multi-photon resonances.
In the intense-pumping regime $F\gg U$, instead, these resonances are replaced by a continuous and monotonous increase of the photon density, up to a sharp transition from a high- to a low-density phase. 
Corresponding to the drop in $\corr{}{}$, the normalized second-order correlation function $g^{(2)}$ exhibits a sharp peak, shown in the bottom panel of Fig.~\ref{Fig:BreakdownG0}.
This quantity is defined as \mbox{$g^{(2)}=\corr{2}{2}/\corr{}{}^2$}.
At the transition, photons have a highly super-Poissonian distribution ($g^{(2)}\gg1$).

\begin{figure}[t]
	\includegraphics[width=0.45\textwidth]{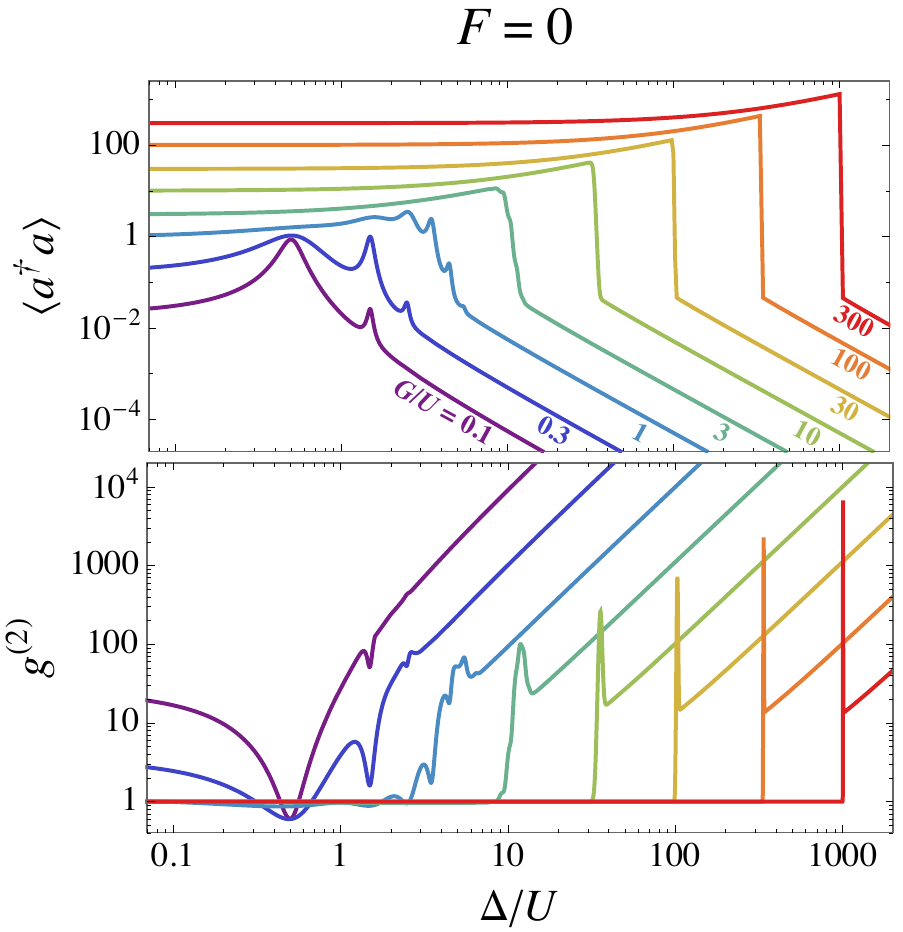}
	\caption{(Color online)
		Same as Fig.~\ref{Fig:BreakdownG0}, but in the presence of two-photon driving only, i.e. $F=0$ and $G\neq0$.	Different curves (and colors) correspond to different pumping strengths $G/U$, spanning from 0.1 to 300 as labeled beside each curve in the top panel.
		One- and two-photon dissipation rates were set to $\gamma=\eta=0.03U$.}
	\label{Fig:BreakdownF0}
\end{figure}

A similar analysis can be performed in the presence of two-photon pumping.
The results obtained for $F=0$ and different values of $G/U$ are presented in Fig.~\ref{Fig:BreakdownF0}.
In the top panel, we observe a similar behavior of the photon density as in Fig.~\ref{Fig:BreakdownG0}.
Note that the analytic solution allows us to reach very high values of $\corr{}{}$ (up to $\sim1300$ photons for $G=300U$).
The behavior of the second-order correlation function $g^{(2)}$ dramatically differs from the case $G=0$ considered in Fig.~\ref{Fig:BreakdownG0}.
For $G\geqslant10U$, we find a sub-Poissonian statistics ($g^{(2)}<1$) for small $\Delta$ and a strong peak corresponding to the drop in density.
After the peak, in the low-density phase, $g^{(2)}$ drops but stays considerably larger than one and, furthermore, it keeps growing roughly quadratically.
This high probability of observing photons pairwise is a clear consequence of the two-photon pumping mechanism.

\begin{figure}[t!]
	\includegraphics[width=0.42\textwidth]{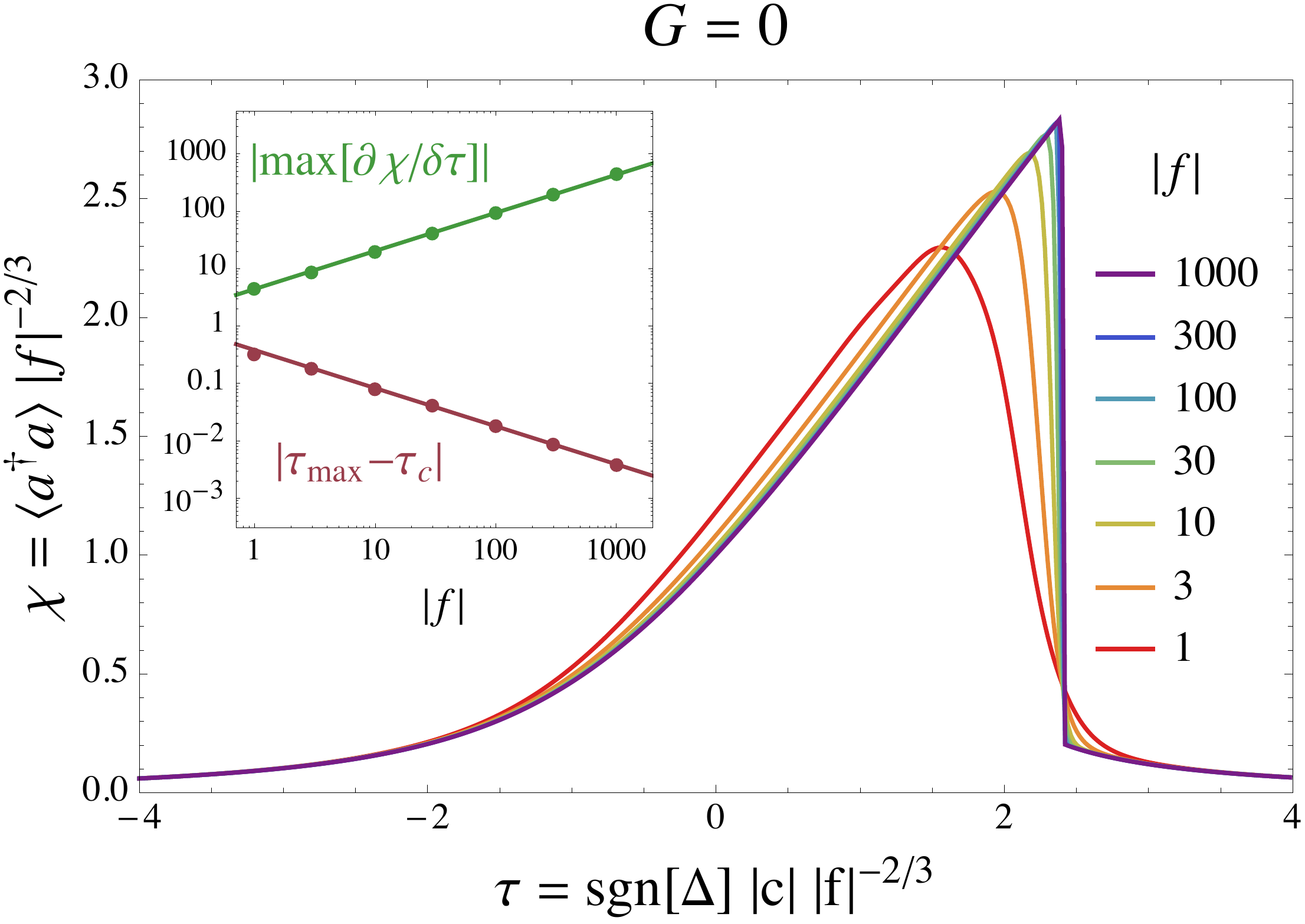}\vspace{0.5cm}
	\includegraphics[width=0.42\textwidth]{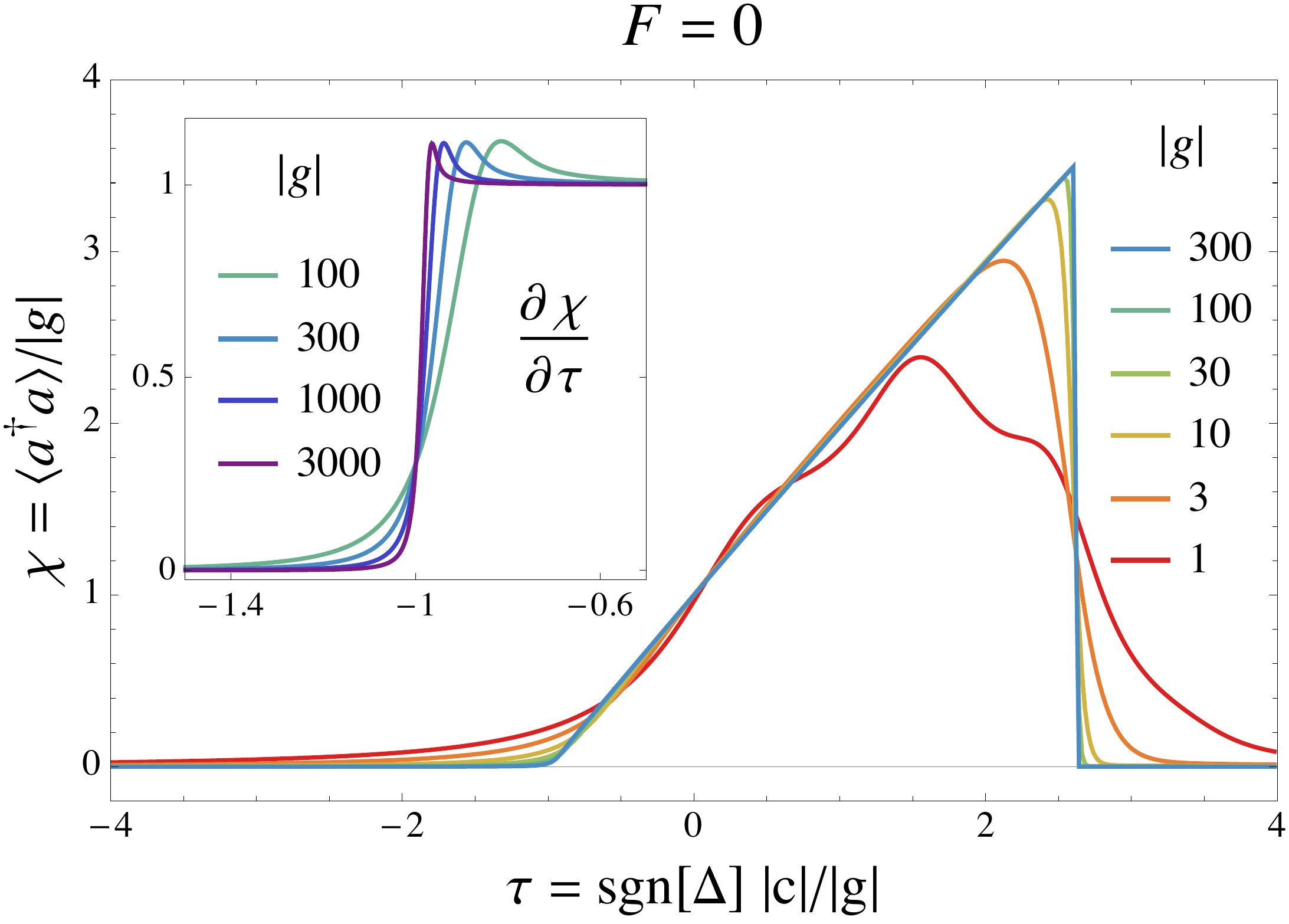}
	\caption{(Color online)
		Top: for the case $G=0$ (coherent driving only), rescaled mean photon density $\chi=\corr{}{}/|f|^{2/3}$ as a function of the dimensionless parameter $\tau={\rm sgn}[\Delta]|c|/|f|^{2/3}$ for different values of the dimensionless coherent drive intensity $|f|$ (see legend).
		The smoothest curve corresponds to $|f|=1$, while for increasing values of $|f|$ the curve gets steeper, acquiring a triangular shape.
		The points in the inset mark height and position of the peak in $\partial\chi/\partial\tau$ as a function of $|f|$.
		The solid lines are power-law fits with exponents $\pm2/3$, performed on the last four points.
		Bottom: for the case $F=0$ (two-photon driving only), rescaled mean photon density $\chi=\corr{}{}/|g|$ as a function of $\tau={\rm sgn}[\Delta]|c|/|g|$ for different values of $|g|$ (see legend).
		The smoothest curve corresponds to $|g|=1$, while $\chi$ progressively tends to a triangular-shaped curve for increasing $|g|$.
		In the inset we show the rapid growth in the derivative $\partial\chi/\partial\tau$ around $\tau=-1$. The smoothest behavior corresponds to $|g|=100$ and the curve progressively acquires a discontinuity by increasing $|g|$ (cf. inset legend).
		Overall, dissipations have been set to $\eta=0.1U$ and $\gamma=0.1|\Delta|$.}
	\label{Fig:PhaseTransition}
\end{figure}

The abrupt change in the density observed above both for $G=0$ and $F=0$ is the result of a dissipative phase transition.
This kind of phenomenon arises in the nonequilibrium context due to the interplay of nonlinearity, drive and dissipation~\cite{AlsingQO91,KilinJOSAB91,CarmichaelPRX15}.
Hence, a dissipative phase transition similar to the one studied numerically by Carmichael for the Jaynes-Cummings model~\cite{CarmichaelPRX15} is also present in our Kerr system.
Our exact solutions proves it unambiguously and allows us to capture also the critical exponents.
In order to further characterize the transition, we consider a scaling which leads to an universal behavior in the thermodynamic limit.
In the coherent-pumping case $G=0$, starting from the semiclassical equation~\eqref{Eq:GP}, one finds that for a large photon number $\corr{}{}\propto|f|^{2/3}$ [as a reminder, $f=F/(U-i\eta)$].
Hence, we expect an universal behavior of the quantity $\chi=\corr{}{}|f|^{-2/3}$.
In Fig.~\ref{Fig:BreakdownG0}, we saw that the high-to-low density transition is triggered by the detuning $\Delta$.
In a more general description, we expect the phase transition to be controlled by the dimensionless complex detuning $c=(\Delta+i\gamma/2)/(U-i\eta)$.
Hence, in the top panel of Fig.~\ref{Fig:PhaseTransition} we show the behavior of $\chi(\tau)$ for $\tau={\rm sgn}[\Delta]|c||f|^{-2/3}$.
In the thermodynamic limit $|f|\to\infty$, $\chi(\tau)$ shows a discontinuous first-order phase transition.
For finite values of $|f|$, the derivative $\partial\chi/\partial\tau$ is peaked at the transition point.
We find that the height and position of this peak follow the power-law behaviors $\max[\partial\chi/\partial\tau]\propto|f|^{2/3}$ and $|\tau_{\rm max}-\tau_c|\propto|f|^{-2/3}$ [cf. inset of Fig.~\ref{Fig:PhaseTransition}~(top)].
For the selected parameters, we find $\tau_c\sim2.41$.

We now perform the same analysis for the two-photon driven case $F=0$, for which, in the thermodynamic limit, one expects $\corr{}{}\propto|g|$ [with $g=G/(u-i\eta)$].
In the bottom panel of Fig.~\ref{Fig:PhaseTransition} we plot, for different values of $|g|$, the function $\chi(\tau)$ where we defined $\chi=\corr{}{}/|g|$ and $\tau={\rm sgn}[\Delta]|c|/|g|$.
Once again, the behavior becomes universal for $|g|\ggg1$, with a sharp transition at positive detuning.
The critical-exponent analysis of the derivative is compatible with $\max[\partial\chi/\partial\tau]\propto|g|$ and $|\tau_{\rm max}-\tau_c|\propto1/|g|$ for $\tau_c\sim2.62$.
The divergent behavior of the derivative in the thermodynamic limit signals the first-order nature of this phase transition.
The latter has the same character of the one observed for $G=0$ and both occur in the regime for which the semiclassical solution predicts optical multistability.
Remarkably, in the case $F=0$ we can identify another phase transition, taking place for $\tau\simeq-1$.
Although $\chi$ stays continuous in the thermodynamic limit, its derivative, shown in the inset of Fig.~\ref{Fig:PhaseTransition}~(bottom), acquires a discontinuity.
This second-order phase transition has no counterpart in the driven-dissipative Kerr model without the two-photon processes.
It takes place around the semiclassical bifurcation point, i.e. when the Wigner function acquires a bi-modal character.

\section{Conclusions and perspectives}\label{sec:conclusion}

In this work, we have investigated the paradigmatic model of a driven-dissipative nonlinear resonator subject to both one- and two-photon processes.
We have shown that, remarkably, the steady-state of such system can be analytically obtained through the complex $P$-representation of its density matrix.
The exact solution, benchmarked against numerical techniques, stands as a powerful tool to investigate the physics of this general model.

We have discussed the limitations of the semiclassical approach in the high-density regime by comparing its prediction to the analytic results.
In this context, we pointed out the emergence of multi-modal Wigner functions, whose structure can not be fully determined semiclassically.
We have also shown that the multimodal character of $W(z)$ can be controlled by external parameters, such as the relative phase of the one- and two-photon pumps.
Furthermore, the exact solution allowed us to explore the physics of a first-order dissipative phase transition in the regime where the semiclassical approach predicts optical multistability.
Moreover, in the two-photon-driven Kerr model (i.e., for $F=0$) we also revealed a second-order phase transition.
The latter has no counterpart in the driven-dissipative Kerr model with $G=0$.

Both theoretical and experimental perspectives of the present work are numerous.
The one- and two-photon driven-dissipative resonator is already realizable with present techniques~\cite{LeghtasScience15}.
The exact solution allows us to predict how the external experimental parameters affects the steady state. Hence, one can generate and manipulate precisely coherent-like states or superpositions of them, which is of great interest towards quantum computation~\cite{GilchristJOB04,OurjoumtsevScience06,MirrahimiNJP14,GotoPRA16,PuriarXiv16}.
The exact results of this work, combined with mean-field~\cite{LeBoitePRL13,LeBoitePRA14} and renormalization techniques~\cite{SchollwockRMP05,FinazziPRL15}, pave the way to the study of exotic many-body phases of light in networks of nonlinear resonators.
Indeed, the flourishing field of reservoir and coupling engineering in circuit QED makes it possible to envision a plethora of combinations between one- and two-photon driving, dissipation, and hopping mechanisms~\cite{JinPRL13,JinPRA14}.
Moreover, effective two-photon processes can arise in the momentum-space Hamiltonian of systems that do not include two-photon mechanisms in real space.
For example, this is the case for a single-cavity polarization-dependent corss-Kerr model~\cite{ParaisoNatMat10,TakemuraNatPhys14} and for the driven-dissipative Bose-Hubbard model~\cite{GeraceNatPhys09,CarusottoPRL09,CarusottoRMP13,WilsonarXiv16}.

\section*{Acknowledgment}
We acknowledge support from the ERC via the Consolidator Grant “CORPHO”, No.~616233.


\appendix
\section{Detailed derivation of the steady-state solution}\label{App:Solution}

In this appendix we provide details about the analytic solution for the one- and two-photon driven-dissipative resonator, whose master equation for the density matrix $\hat{\rho}$ is~\eqref{Eq:MasterEquationOur}.

\subsection{From the master equation to the Fokker-Planck}\label{subsec:solFP}

As stated in the main text, the problem is tackled by writing $\hat{\rho}$ in terms of the complex $P$-representation~\eqref{Eq:PRepresentation}.
Although different choices for the $P$-representation are possible~\cite{DrummondJPA80b}, the complex one is the best candidate to find an exact solution for the considered class of driven-dissipative problems~\cite{DrummondJPA80a}.
The advantage of this approach is that the master equation for $\hat{\rho}$ can be translated into a partial differential equation for the function $P(\alpha,\beta)$~\cite{DrummondJPA80b,WallsBOOK}.
Indeed, the action of the annihilation and creation operators on the projector $\ket{\alpha}\bra{\beta^*}$ establishes a term-by-term conversion between elements of the master equation and differential operators.
As an example, consider $\aop\hat{\rho}$: from Eq.~\eqref{Eq:PRepresentation} one sees that the action of $\aop$ gives a multiplication of the integrand by $\alpha$.
Similarly, we can get all the following translation rules:
\begin{subequations}\label{Eq:MasterToFP}
	\begin{align}
	\aop\hat{\rho}\leftrightarrow&
	\alpha\, P(\alpha,\beta),
	\\
	\ad\hat{\rho}\leftrightarrow&
	\left(\beta-\partial_\alpha\right) P(\alpha,\beta),
	\\
	\hat{\rho}\aop\leftrightarrow&
	\left(\alpha-\partial_\beta\right) P(\alpha,\beta),
	\\
	\hat{\rho}\ad\leftrightarrow&
	\beta\, P(\alpha,\beta).
	\end{align}
\end{subequations}
Exploiting~\eqref{Eq:PRepresentation} and~\eqref{Eq:MasterToFP}, matching the terms inside the integrals, one gets that the function $P(\alpha,\beta)$ must satisfy the Fokker-Planck-like equation
\begin{equation}\label{Eq:FokkerPlanck}
 i \,\partial_t P=
\sum_{i=\alpha,\beta} \partial_i \left[-A^i P
+ \frac{1}{2} \sum_{j=\alpha,\beta} \partial_j \left(D^{ij} P \right)  \right],
\end{equation}
where $A^i$ indicates the components of the drift vector
\begin{equation}\label{Eq:DriftA}
\bar{A}=
\begin{pmatrix}
-\widetilde{\Delta} \alpha + \widetilde{U}^* \alpha^2 \beta + F + G \beta \\
\widetilde{\Delta}^* \beta - \widetilde{U} \alpha \beta^2 -F^* -G^* \alpha
\end{pmatrix},
\end{equation}
and $D^{ij}$ is a matrix element of the diffusion tensor
\begin{equation}\label{Eq:DiffusionD}
\bar{\bar{D}}=
\begin{pmatrix}
\widetilde{U}^* \alpha^2 + G & 0 \\
0 & -\widetilde{U} \beta^2 -G^*
\end{pmatrix}.
\end{equation}
In Eqs.~\eqref{Eq:DriftA} and~\eqref{Eq:DiffusionD}, we introduced the complex detuning $\widetilde{\Delta}=\Delta+ i \gamma/2$ and the complex interaction energy $\widetilde{U}=U+ i \eta$.

Being interested in the steady-state density matrix, we seek for the steady-state solution of the Fokker-Planck equation~\eqref{Eq:FokkerPlanck}, i.e. we look for the function $P$ satisfying $\partial_t P=0$. 
Solving the resulting differential equation is generally a hard task.
One can simplify the problem by requiring that every term of the sum vanishes:
\begin{equation}\label{Eq:FPManipulation1}
A^i P
- \frac{1}{2} \sum_{j=\alpha,\beta} \partial_j \left(D^{ij} P \right)=0,
\qquad i=\alpha,\beta.
\end{equation}
After some straightforward algebraic manipulation, Eq.~\eqref{Eq:FPManipulation1} can be cast as
\begin{equation}\label{Eq:FPManipulation2}
2A^i - \sum_{j=\alpha,\beta} \left(\partial_j\,D^{ij}\right)=
\sum_{j=\alpha,\beta} D^{ij}\, \partial_j \ln(P),
\quad i=\alpha,\beta,
\end{equation}
which, if the matrix $\bar{\bar{D}}$ is invertible, is solved by
\begin{equation}\label{Eq:FPManipulation3}
\partial_j \ln(P) = \sum_{i=\alpha,\beta} \left(D^{-1}\right)^{ji}
\left[2A^i - \sum_{k=\alpha,\beta} \left(\partial_k\,D^{ik}\right)\right],
\end{equation}
for $j=\alpha,\beta$.
Hence, we can write $P=\exp(-\phi)$ and treat $\phi$ as a scalar potential in the complex variables $\alpha$ and $\beta$.
Such a potential defines a generalized force $\bar{\Phi}=-\bar{\nabla}\phi$ of components
\begin{equation}\label{Eq:GeneralizedForce}
\Phi_j=-\partial_j\,\phi=\sum_{i=\alpha,\beta}
\left(D^{-1}\right)^{ji}
\left[2A^i - \sum_{k=\alpha,\beta} \left(\partial_k\,D^{ik}\right)\right].
\end{equation}
For the function $\phi$ to be a well-behaved potential, one must require that the crossed derivatives of the force components~\eqref{Eq:GeneralizedForce} are the same, that is
\begin{equation}\label{Eq:PotentialConditions}
\partial_i \Phi_j=\partial_j \Phi_i.
\end{equation}
The latter are known as the potential conditions.
They also ensure that the integral of the coupled differential equations $\partial_j\,\phi=-\Phi_j$ ($j=\alpha,\beta$) is independent of the integration path.
Hence, it is possible to obtain $\phi$ as
\begin{equation}\label{Eq:GeneralPotentialIntegral}
\phi(\alpha,\beta)=\phi(\alpha_0,\beta_0)-\int_\Gamma \bar{\Phi}(\alpha',\beta')\cdot d\bar{s}(\alpha',\beta'),
\end{equation}
where $d\bar{s}(\alpha',\beta')$ is an infinitesimal displacement element along the path $\Gamma$ going from the arbitrary reference point $\{\alpha_0,\beta_0\}$ to $\{\alpha,\beta\}$.

Let us now consider our specific case.
Starting from the definitions of $\bar{A}$ and $\bar{\bar{D}}$ given in Eqs.~\eqref{Eq:DriftA} and~\eqref{Eq:DiffusionD}, we find the force
\begin{equation}\label{Eq:GeneralizedForceOur}
\bar{\Phi}=2
\begin{pmatrix}
\frac{F+G\beta-\left(\widetilde{\Delta}+\widetilde{U}^*\right)\alpha
	+\widetilde{U}^*\alpha^2\beta}
{G+\widetilde{U}^*\alpha^2} \\[8pt]
\frac{F^*+G^*\alpha-\left(\widetilde{\Delta}^*+\widetilde{U}\right)\beta
	+\widetilde{U}\beta^2\alpha}
{G^*+\widetilde{U}\beta^2}
\end{pmatrix},
\end{equation}
which fulfills the potential conditions~\eqref{Eq:PotentialConditions}.
\begin{widetext}
	To get the corresponding potential, we use Eq.~\eqref{Eq:GeneralPotentialIntegral} for the path $\Gamma:=\{0,0\}\to\{\alpha,0\}\to\{\alpha,\beta\}$, which formally gives
	\begin{equation}
	\phi(\alpha,\beta)=\phi(0,0)
	-\int_{\{0,0\}}^{\{\alpha,0\}} \Phi_\alpha(\alpha',0)\, d\alpha'
	-\int_{\{\alpha,0\}}^{\{\alpha,\beta\}} \Phi_\beta(\alpha,\beta')\, d\beta'.
	\end{equation}
	Performing the integration and discarding irrelevant constant terms, we get
	\begin{equation}\label{Eq:PotentialOur}
	\phi(\alpha,\beta)=
	\ln\left[\left(\alpha^2+g\right)^{1+c}\left(\beta^2+g^*\right)^{1+c^*}\right]
	-2\alpha\beta
	+\frac{2f}{\sqrt{g}}\arctan\left(\frac{\sqrt{g}}{\alpha}\right)
	+\frac{2f^*}{\sqrt{g^*}}\arctan\left(\frac{\sqrt{g^*}}{\beta}\right),
	\end{equation}
where we introduced the dimensionless quantities $c=\widetilde{\Delta}/\widetilde{U}^*$, $f=F/\widetilde{U}^*$, and $g=G/\widetilde{U}^*$.
It follows immediately that the $P$-representation of the density matrix for the one- and two-photon driven-dissipative resonator is the one given in Eq.~\eqref{Eq:PRepOur}.
As a final check, one can easily verify that the $P(\alpha,\beta)$ given in Eq.~\eqref{Eq:PRepOur} solves the Fokker-Planck-like equation~\eqref{Eq:FokkerPlanck} for the steady-state regime. 

\end{widetext}

\section{Analytic integration of the $\FF$ functions}\label{App:IntegrationFF}

\begin{figure}[t!]
	\includegraphics[width=0.48\textwidth]{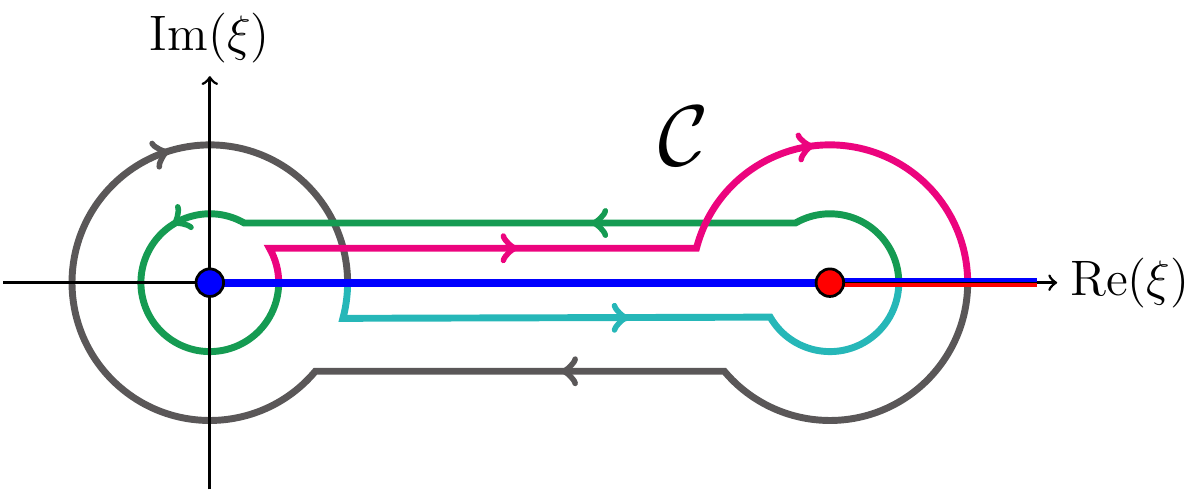}
	\caption{(Color online)
		Representation of the Pochhammer path on the complex plane $\{\Re(\xi),\Im(\xi)\}$.
		The blue and red dot represent the poles of the integrand in Eq.~\eqref{Eq:Pochhammer}, located at $\xi=0,1$ together with the corresponding branch cuts (form each pole towards $+\infty$ on the real axis).
		The Pochhammer contour $\mathcal{C}$ crosses each of the cuts an equal number of times in one direction and in the opposite, hence, for any starting point, it begins and ends on the same Riemann sheet.
		The change in the path color emphasizes the passage to a different Riemann surface.}
	\label{fig:Poch}
\end{figure}

In this appendix we detail the analytic integration of the $\FF_m$ functions defined in Eq.~\eqref{Eq:DefinitionFFormal} and in terms of which we expressed the analytic steady-state solutions presented in Sec.~\ref{Sec:Solution}.
Let us start by using the identity $2\, i \arctan(z)=\ln(1+ i  z)-\ln(1- i  z)$ in Eq.~\eqref{Eq:DefinitionFFormal} to write
\begin{equation}
\FF_m\left(f,g,c\right)=\!\!\int_\mathcal{\!C} d\alpha\,\alpha^m
\frac{\left(\alpha+ i \sqrt{g}\right)^{\varphi-(1+c)}}
{\left(\alpha- i \sqrt{g}\right)^{\varphi+(1+c)}},
\end{equation}
where we also introduced $\varphi= i  f/\sqrt{g}$.
One can now perform a change of variable, keeping in mind that $\mathcal{C}$ must be a closed path encircling all the singularities of the integrand.
Hence, we chose $\alpha= i \sqrt{g}(1-2\xi)$, which gives
\begin{align}
\FF_m&\left(f,g,c\right)=
\frac{( i \sqrt{g})^m}{(-1)^{\varphi+c}\,(2 i \sqrt{g})^{1+2c}}
\nonumber\\
&\times\int_\mathcal{\!C} d\xi\,\xi^{-\varphi-c-1}\,
(1-\xi)^{\varphi-c-1}\,(1-2\xi)^m.
\end{align}
We are left with a complex-plane integral of the form
\begin{align}\label{Eq:Pochhammer}
\mathcal{I}&(\eta;c_1,c_2)=\int_\mathcal{\!C} d\xi\,
\xi^{c_1-1} (1-\xi)^{c_2-1} (1-\eta\,\xi)^m
\nonumber\\
&=\sum_{k=0}^{m} \left(-\eta\right)^k\, \binom{m}{k}
\int_\mathcal{\!C} d\xi\, \xi^{c_1+k+1} (1-\xi)^{c_2-1}.
\end{align}
The path $\mathcal{C}$ must encircle both the poles at $\xi=0$ and $\xi=1$.
Furthermore, for $\mathcal{C}$ to be properly closed one must take into account the presence of two branch cuts going from each pole to $|\xi|\to\infty$.
A convenient choice is the Pochhammer contour~\cite{MacRobertBOOK}, which is sketched in Fig.~\ref{fig:Poch}.
Such a path correctly encircles the poles and crosses the branch cuts an equal number of times in one sense and in the opposite one (a property which does not depend on the cuts orientation).
Hence, the path is closed since it begins and ends on the same Riemann sheet.
After integration along the Pochhammer path, one gets
\begin{align}
&\mathcal{I}[\eta;\alpha,\beta]
=\left(1-e^{2\pi ic_1}\right)\!\left(1-e^{2\pi ic_2}\right)
\nonumber\\&\qquad\times
\frac{\Gamma\left(c_1\right)\,\Gamma\left(c_2\right)}
{\Gamma\left(c_1+c_2\right)}
\,_2F_1\left(-m,c_1;c_1+c_2;\eta\right),
\end{align}
being $\,_2 F_1$ the Gauss hypergeometric function
\begin{align}
\,_2F_1&\left(-m,c_1;c_1+c_2;\eta\right)=\nonumber\\
&\sum_{k=0}^{m} \left(-\eta\right)^k\, \binom{m}{k}
\frac{\Gamma\left(k+c_1\right)\,\Gamma\left(c_1+c_2\right)}
{\Gamma\left(k+c_1+c_2\right)\,\Gamma\left(c_1\right)}.
\end{align}
Finally, we obtain Eq.~\eqref{Eq:FormulaFcomplete},
where all the $m$-independent prefactors have been dropped.
Indeed, in the calculation of all physical quantities (cf. Sec.~\ref{subsec:ssvalues}) those terms, depending only on the parameters $f$, $g$, and $c$, are canceled by the normalization coefficient $\NN$.

\section{Series convergence and closed forms}

\begin{figure}[t]
	\includegraphics[width=0.46\textwidth]{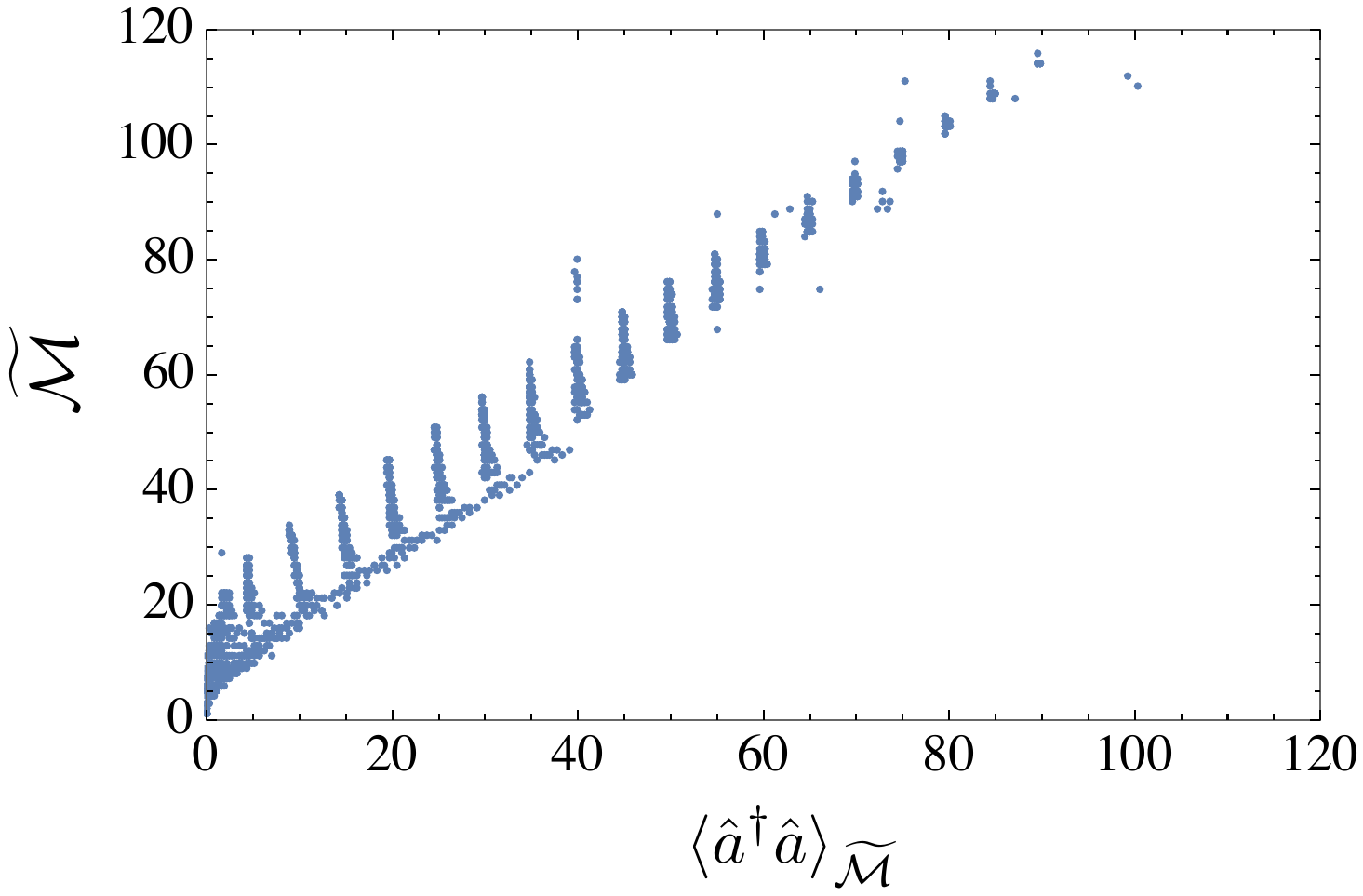}
	\caption{Numerical cutoff $\widetilde{\mathcal{M}}$ as a function of the corresponding mean photon density $\corr{}{}_{\widetilde{\mathcal{M}}}$ [Eq.~\eqref{Eq:MeanPhotPartial}] for different system parameters.
		The plot has been obtained setting the convergence criterion as explained in the main text.
		Each point of the diagram corresponds to $\gamma=\eta=0.1 U$, while we varied $\Delta/U\in[-30,30]$ and $F/U,G/U\in[0,60]$}
	\label{Fig:Convergence}
\end{figure}

The analytic results for the one- and two-photon driven-dissipative nonlinear resonator given in Eqs.~\eqref{Eq:Normalization}, \eqref{Eq:CorrijGeneral}, \eqref{Eq:RhopqGeneral}, and~\eqref{Eq:WignerGeneral}, contain infinite summations which, in the general case $F,G\neq0$ must be estimated numerically.
In this appendix we take as an example the photon number $\corr{}{}$ to show that such series converge fast in a wide range of parameters.
Moreover, we give the exact closed forms of $\corr{i}{j}$, $\rho_{pq}$, and $W(z)$ for the case $F=0$.

\subsection{Convergence of the series in the general case}\label{Subsec:Convergence}

To investigate the convergence of the series defined by Eq.~\eqref{Eq:CorrijGeneral}, let us consider the mean photon number $\braket{\hat{a}^\dagger\hat{a}}$.
In this expression there are two sums to evaluate: the one explicitly expressed in Eq.~\eqref{Eq:CorrijGeneral} and a second one appearing in the normalization $\NN$ [Eq.~\eqref{Eq:Normalization}].
The convergence of $\braket{\hat{a}^\dagger\hat{a}}$ can be examined in terms of a single parameter $\mathcal{M}$, the cutoff of both sums.
Hence, we introduce
\begin{equation}\label{Eq:MeanPhotPartial}
\braket{\hat{a}^\dagger\hat{a}}_\mathcal{M}=
\frac{\sum_{m=0}^\mathcal{M} \frac{2^m}{m!}  \left|\mathcal{F}_{m+1}(f,g,c)\right|^2}
{\sum_{m=0}^\mathcal{M} \frac{2^m}{m!}  \left|\mathcal{F}_m(f,g,c)\right|^2}.
\end{equation}
In the present work, we controlled the convergence by verifying that the addition of two further elements does not affect the result beyond the 6$^{\rm th}$ relevant digit.
In other words, we chose the smallest $\widetilde{\mathcal{M}}$ ensuring that
$\left|
\braket{\hat{a}^\dagger\hat{a}}_{\widetilde{\mathcal{M}}}-
\braket{\hat{a}^\dagger\hat{a}}_{\widetilde{\mathcal{M}}-2}
\right|<10^{-6}\braket{\hat{a}^\dagger\hat{a}}_{\widetilde{\mathcal{M}}}$.
In Fig.~\ref{Fig:Convergence}, we show the required cutoff $\widetilde{\mathcal{M}}$ as a function of the corresponding value of the mean photon number for different system parameters.
It turns out that $\widetilde{\mathcal{M}}$ grows roughly linearly with $\corr{}{}_{\widetilde{\mathcal{M}}}$.
We verified that similar convergence criteria efficiently applies to the other quantities defined by Eqs.~\eqref{Eq:CorrijGeneral}, \eqref{Eq:RhopqGeneral}, and~\eqref{Eq:WignerGeneral}, in a wide range of system parameters.
In general, the numerical evaluation of the exact solution can be performed with arbitrary precision.
Such a computation is faster and much less memory demanding than numerical approaches, in particular for high-density regimes.

\subsection{Closed forms for $F=0$}\label{subsec:closedForms}
The general model described by the master equation~\eqref{Eq:MasterEquationOur} can be specialized to many different cases.
Among them, a case of particular interest is the one in which the one-photon pumping is absent, that is when $F=0$~\cite{KryuchkyanOC96,MeaneyEPJQT14,LeghtasScience15,MingantiSciRep16}.
Remarkably, in this case one finds that
\begin{subequations}\label{Eq:FSimplified}
	\begin{align}
	\FF_{2m+1}\left(0,g,c\right)&=0,
	\\
	\FF_{2m}\left(0,g,c\right)&=\left(i\sqrt{g}\right)^{2m}
	\,_2\widetilde{F}_1\left(-2m,-c;-2c;2\right)
	\nonumber\\
	&=(-g)^m
	\frac{1}{\sqrt{\pi}}
	\frac{\Gamma\left(\frac{1}{2}-c\right)}
	{\Gamma\left(-2c\right)}\,
	\frac{\Gamma\left(\frac{1}{2}+m\right)}
	{\Gamma\left(\frac{1}{2}+m-c\right)}
	\nonumber\\
	&\equiv
	(-g)^m
	\frac{\Gamma\left(\frac{1}{2}+m\right)}
	{\Gamma\left(\frac{1}{2}+m-c\right)},
	\end{align}
\end{subequations}
where, in the last identity, we dropped further $m$-independent factors which would be naturally absorbed in the normalization.

\begin{widetext}
	Making use of Eqs.~\eqref{Eq:FSimplified}, we derived the following closed forms:
	\begin{subequations}\label{Eq:CorrijNoF}
		\begin{align}
		\corr{(2i)}{(2j)}
		&=
		\frac{\Gamma\left(\frac{1}{2}+j\right)\Gamma\left(\frac{1}{2}+i\right)}{\sqrt{\pi}}\,
		(-g)^j(-g^*)^i\,
		\frac{\,_2\widetilde{F}_3\left(\frac{1}{2}+j,\frac{1}{2}+i;
			\frac{1}{2},\frac{1}{2}+j-c,\frac{1}{2}+i-c^*;|g|^2\right)}
		{\,_1\widetilde{F}_2\left(\frac{1}{2};\frac{1}{2}-c,\frac{1}{2}-c^*;|g|^2\right)},
		\\
		\corr{(2i+1)}{(2j+1)}
		&=
		\frac{\Gamma\left(\frac{3}{2}+j\right)\Gamma\left(\frac{3}{2}+i\right)}{\sqrt{\pi}}\,
		(-g)^{j+1}(-g^*)^{i+1}\,
		\frac{\,_2\widetilde{F}_3\left(\frac{3}{2}+j,\frac{3}{2}+i;
			\frac{3}{2},\frac{3}{2}+j-c,\frac{3}{2}+i-c^*;|g|^2\right)}
		{\,_1\widetilde{F}_2\left(\frac{1}{2};\frac{1}{2}-c,\frac{1}{2}-c^*;|g|^2\right)},
		\\
		\corr{(2i)}{(2j+1)}&=\corr{(2i+1)}{(2j)}=0,
		\end{align}
	\end{subequations}
	\begin{subequations}\label{Eq:RhopqNoF}
		\begin{align}
		\braket{2p|\hat{\rho}|2q}
		&=\frac{\Gamma\left(\frac{1}{2}+p\right)\Gamma\left(\frac{1}{2}+q\right)}{\sqrt{\pi(2p)!(2q)!}}\,
		(-g)^p(-g^*)^q\,
		\frac{\,_2\widetilde{F}_3\left(\frac{1}{2}+p,\frac{1}{2}+q;
			\frac{1}{2},\frac{1}{2}+p-c,\frac{1}{2}+q-c^*;\left|\frac{g}{2}\right|^2\right)}
		{\,_1\widetilde{F}_2\left(\frac{1}{2};\frac{1}{2}-c,\frac{1}{2}-c^*;|g|^2\right)},
		\\
		\braket{2p+1|\hat{\rho}|2q+1}
		&=\frac{\Gamma\left(\frac{3}{2}+p\right)\Gamma\left(\frac{3}{2}+q\right)}{2\sqrt{\pi(2p+1)!(2q+1)!}}\,
		(-g)^{p+1}(-g^*)^{q+1}\,
		\frac{\,_2\widetilde{F}_3\left(\frac{3}{2}+p,\frac{3}{2}+q;
			\frac{3}{2},\frac{3}{2}+p-c,\frac{3}{2}+q-c^*;\left|\frac{g}{2}\right|^2\right)}
		{\,_1\widetilde{F}_2\left(\frac{1}{2};\frac{1}{2}-c,\frac{1}{2}-c^*;|g|^2\right)},
		\\
		\braket{2p|\hat{\rho}|2q+1}&=\braket{2p+1|\hat{\rho}|2q}=0,
		\end{align}
	\end{subequations}
\end{widetext}

\begin{equation}
\label{Eq:WignerNoF}
W(z)=\frac{2}{\pi}\,
\frac{\left|
	\,_0{F}_1\left[\frac{1}{2}-c;-g(z^*)^2\right]
	\right|^2}
{{}_1{F}_2\left[\frac{1}{2};\frac{1}{2}-c,\frac{1}{2}-c^*;|g|^2\right]}
\, e^{-2|z|^2}.
\end{equation}
In the equations above, $_p F_q$ denotes a generalized hypergeometric function~\cite{BaileyBOOK}, defined by the analytic extension of
\begin{equation}
\,_p F_q(a_1,\cdots,a_p;\,b_1,\cdots,b_p;\, z)=
\sum_{k=0}^{\infty}\frac
{\prod_{n=1}^{p} \left(a_n\right)_k}
{\prod_{m=1}^{q} \left(b_m\right)_k}
\frac{z^k}{k!},
\end{equation}
with \mbox{$\left(a\right)_k=\Gamma(a+k)/\Gamma(a)$}.
The regularized hypergeometric functions $_p \widetilde{F}_q$ are defined as
\begin{align}
\,_p \widetilde{F}_q&(a_1,\cdots,a_p;\,b_1,\cdots,b_p;\, z)=
\nonumber\\
&\frac{\,_p F_q(a_1,\cdots,a_p;\,b_1,\cdots,b_p;\, z)}
{\prod_{m=1}^{q} \Gamma(b_m)}.
\end{align}
The closed forms presented above are computationally much more efficient than the corresponding implicit forms~\eqref{Eq:CorrijGeneral}, \eqref{Eq:RhopqGeneral}, and~\eqref{Eq:WignerGeneral}.



%

\end{document}